\documentclass[sigconf]{acmart}
\AtBeginDocument{%
  }

\usepackage{tikz}
\usetikzlibrary{positioning}

\begin{document}

\title{Literature Review Of Multi-Agent Debate For Problem-Solving}

\author{Arne Tillmann}
\email{a.tillmann@stud.uni-goettingen.de}
\affiliation{
  \institution{University of Göttingen}
  \city{Göttingen}
  \state{Lower Saxony}
  \country{Germany}
}

\renewcommand{\shortauthors}{Arne Tillmann}

\begin{abstract}
Multi-agent large language models (MA-LLMs) are a rapidly growing research area that leverages multiple interacting language agents to tackle complex tasks, outperforming single-agent large language models. This literature review synthesizes the latest research on agent profiles, communication structures, and decision-making processes, drawing insights from both traditional multi-agent systems and state-of-the-art MA-LLM studies. In doing so, it aims to address the lack of direct comparisons in the field, illustrating how factors like scalability, communication structure, and decision-making processes influence MA-LLM performance. By examining frequent practices and outlining current challenges, the review reveals that multi-agent approaches can yield superior results but also face elevated computational costs and under-explored challenges unique to MA-LLM. Overall, these findings provide researchers and practitioners with a roadmap for developing robust and efficient multi-agent AI solutions.

\end{abstract}


\maketitle

\section{Introduction}
\label{sec-introduction}

The reference to the latest OpenAI model serves as evidence for increasingly powerful Large Language Models (LLMs) \citep{HttpsCdnopenaicomGpt45systemcard2272025pdf}. In a sense, we are running out of benchmarks. The latest benchmarks built required domain experts to be constructed \cite{glazerFrontierMathBenchmarkEvaluating2024, reinGPQAGraduateLevelGoogleProof2023}.

Increase in compute \citep{EpochNotableModels2024}, algorithmic efficacy \cite{erdilAlgorithmicProgressComputer2023}, and algorithmic progress \citep{post-training} drive the improvements. Compute refers to the processing power, memory, and storage used for training and inference \citep{IntroductionSITUATIONALAWARENESS}. Ways to improve algorithmic efficiency are better optimization methods, architectural innovations, or techniques that reduce resource requirements. By algorithmic progress, I mean innovation that goes beyond just training better base models, such as reinforcement learning from human feedback \citep{christiano2023deepreinforcementlearninghuman}, Chain-of-Thought \citep{wei2023chainofthoughtpromptingelicitsreasoning}, or the usage of tools (web browser, Linux terminal, python interpreter).

The story is not told that easily when it comes to MA-LLMs. What we see is that multi-agent systems often outperform single-agent systems \citet{yinExchangeThoughtEnhancingLarge2023, Chan2023ChatEvalTB, Chen2023ReConcileRC, Wu2023LargeLM} on tasks that require arithmetical reasoning, software development, or even text summarization. This improvement is mainly attributed to the multiple perspectives that different agents can take on and the feedback they can give each other \citep{duImprovingFactualityReasoning2023, rasalNavigatingComplexityOrchestrated2024, suzgunMetaPromptingEnhancingLanguage2024, xuTowardsReasoningLargeLanguage2023, yinExchangeThoughtEnhancingLarge2023}. Compared to better quantifiable factors such as compute, algorithmic efficacy, and the obvious improvements of algorithmic innovations, the perspectives agents can take on, and the feedback they can give each other are much harder to measure. What are valuable perspectives? What is good feedback?

As for single-agent LLMs, the MA-LLMs applications extend beyond problem-solving, acing benchmark after benchmark. They simulate complex real-world environments with application in psychology, policy making, sociology, game theory, and gaming \citep{guoLargeLanguageModel2024}. There is even an effort to construct a meta-benchmark to measure LLM agents' ability to improve other agents \citep{meta-benchmark}. The more capable LLMs become and the more complex the problems we want them to solve, the better they need to be at modeling real-world environments.

Given the significant advancements in the study of MA-LLMs, I answer key questions that remain unanswered in the existing literature on MA-LLMs, particularly regarding their safety and their scaling: 

\paragraph{\textbf{Research Questions:}}
\begin{itemize} \label{research-questions} 
    \item How do MA-LLMs terminate discussions and make decisions? What are the strengths and limitations of established methods?
    \item How do multi-agent LLM systems scale with the number of agents and rounds of debate with respect to their problem-solving performance? What are the challenges arising from scaling MA-LLMs?
\end{itemize}

The rest of the paper is structured as follows: Section~\ref{sec-RelatedLiteratureReviews} reviews recent literature reviews on MA-LLMs and positions our work within the broader context. Section~\ref{sec-methodology} details the methodology used to search, select, and analyze relevant papers. I then describe the general setup of MA-LLM systems for problem-solving, including core agent profiles and communication structures, in Section~\ref{sec-general-setup}. 

Next, the core of the paper answers the research questions. Section~\ref{sec-termination} focuses on termination and decision-making mechanisms, while Section~\ref{sec-scaling} discusses the effects of scaling the number of agents and communication rounds. In Section~\ref{sec-discussion}, I provide a discussion of our main findings and revisit the research questions. Finally, Section~\ref{sec-conclusion} concludes the paper, and Section~\ref{sec-limitations} points to promising directions for future work.

The main contributions of this paper are threefold:
\begin{itemize}
    \item The introduction of a classification scheme from the literature of traditional multi-agent systems to the research field of multi-agent large language models
    \item The systematic description of the most prominent decision-making processes
    \item The analysis of the scaling behavior of MA-LLMs and the identification of the challenges that come with it.
\end{itemize}

\section{Related Literature Reviews}
\label{sec-RelatedLiteratureReviews}

The most comprehensive review of MA-LLM systems to date is presented by \citet{guoLargeLanguageModel2024}, encompassing problem-solving capabilities and real-world simulations. Their taxonomy serves as a foundational framework for this study, with clear indications of overlapping classifications and modifications or additions. Certain classes have been excluded due to the broader scope of their work, allowing for a more focused examination specific to MA-LLM systems for problem-solving. However, several critical research gaps remain unaddressed. While extensive, the survey by \citet{guoLargeLanguageModel2024} is not fully up-to-date with the latest advancements in MA-LLMs analysis, scaling, and learning strategies \citet{ beckerStayFocusedProblem2025, qianScalingLargeLanguageModelbasedMultiAgent2025, chenOptimaOptimizingEffectiveness2025}, and some of their classifications appear arbitrary or overly broad. Moreover, although they acknowledge decision-making processes and scaling as open challenges, they do not delve deeply into these areas.

Other surveys, such as \citet{Multi-Agent-survey}, offer a broader perspective on general multi-agent systems that go beyond MA-LLMs, providing additional categories that enhance our understanding of MA-LLMs in the context of problem-solving. Consequently, like \citet{guoLargeLanguageModel2024}, they lack specificity regarding the unique challenges, decision-making processes and scaling of large language models in multi-agent configurations. Research on scaling laws, as seen in the survey by \citet{li2025misfitting}, provides valuable insights but does not address how MA-LLMs scale with the number of agents or the complexity of tasks.

\section{Methodology}
\label{sec-methodology}
\subsection{Problem-Solving}
\label{sec-problem-solving}

There exist very narrow and extensive definitions for problem-solving. The majority of the analyzed papers in this literature review employ a narrow definition that aligns with the notion presented by \citet{guoLargeLanguageModel2024}. Problem-solving refers to the ability of MA-LLMs or LLMs in general to provide an answer to a task from a benchmark which is then compared to the gold answer. For an extensive list of benchmarks used to evaluate MA-LLM systems consult \citet{guoLargeLanguageModel2024}. However, because the definition is arbitrary, I also include several noteworthy papers in the literature review that suggest how more sophisticated MA-LLM systems address more complex problems not included in widely established benchmarks. 

\subsection{Search with Elicit}
\label{sec-elicit}
To conduct the literature review, I utilize \textbf{Elicit} \cite{elicit}, an AI-powered research assistant designed to streamline the literature discovery and synthesis process. Elicit leverages language models to extract, summarize, and rank academic articles based on user-defined queries, enabling the efficient identification of key studies and insights.

Using Elicit, I directly search for papers aligned with the pre-specified research questions from Section~\ref{research-questions} in the timeframe from 2023 to February 2025. The platform enables me to specify nuanced search criteria, prioritizing recent publications, influential studies, and articles that cite established benchmarks in the field. One feature that turns out to be particularly useful is the \textit{show more like these} papers, which searches for new papers based on a pre-selection. Elicit's ability to summarize key findings and highlight methodologies helps me quickly assess the relevance and quality of each source.

The structured and systematic nature of Elicit.org significantly enhances the rigor of my review process. It reduces the time spent on manual searches and improves the comprehensiveness of the literature surveyed, ensuring that I capture diverse perspectives across subfields. The platform directly links to publications, maintaining the transparency of the extracted information and enabling me to validate and cross-check findings against the original publications. This methodical approach provides a solid foundation for analyzing the state-of-the-art MA-LLM systems for problem-solving. 

In addition to Elicit, I search for relevant literature on Google Scholar, Semantic Scholar, and Google using the following keywords and prompts constructed out of those: \textit{literature review, meta-study, multi-agent, LLM, benchmarks, scaling, termination, decision-making}. I use Citation Gecko to map the citation network of my research area by visualizing the relationships between papers. It helps me discover highly cited papers and uncover connections between works, giving me a comprehensive view of influential research in my field.

\subsection{Reading}
\label{sec-reading}

Eventually, I read the selected papers myself, following a three-step approach. After reading the title and skimming the abstract, the goal of the \textbf{First Impression} is to decide if the paper is relevant. In Elicit, I benchmark those papers that help with the search for more papers like these. Then the \textbf{Bird's Eye View} entails reading the more extended summary generated by Elicit, ChatGPT, or the section and subsection headings, as well as the conclusion, and skimming over the references. Its Goal: Decide if the paper is relevant enough for full reading. Eventually, the \textbf{Full Read} will classify the paper into my existing taxonomy or identify new taxa where necessary.

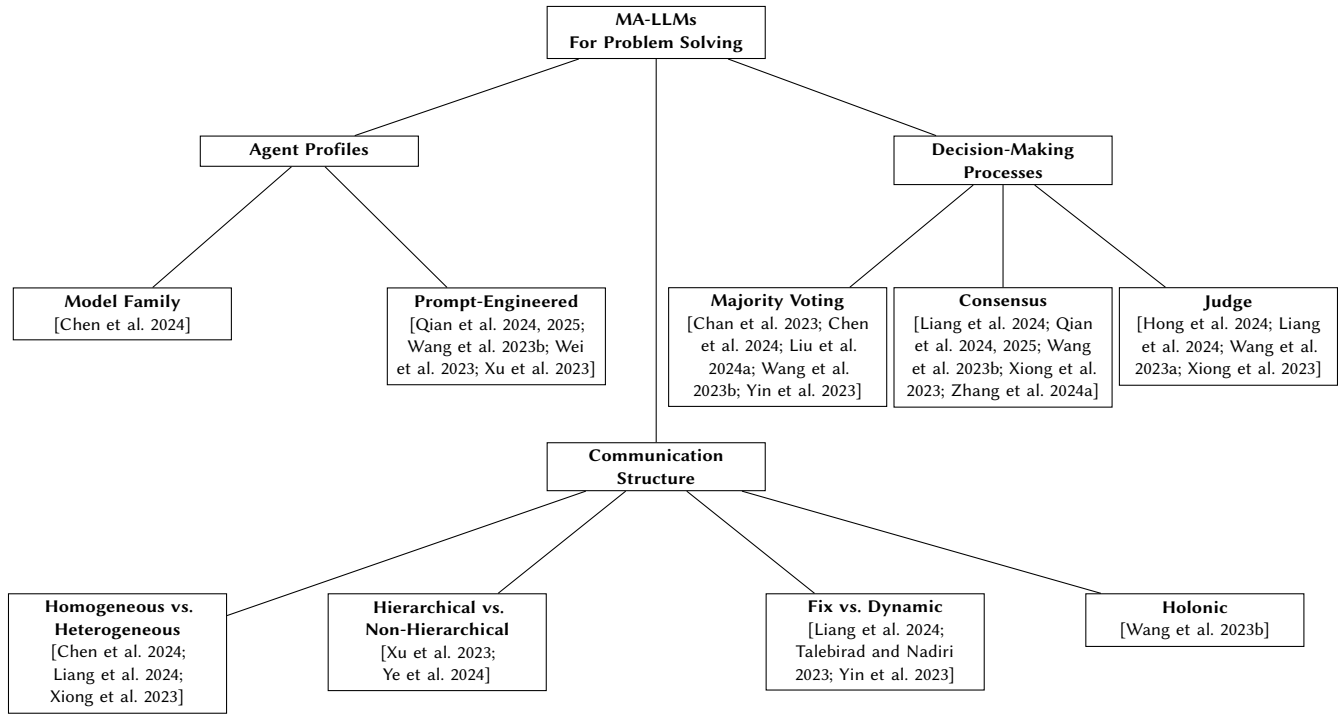
\begin{figure*}
    \hspace{0cm}
\begin{tikzpicture}[
        xshift=-2cm,
        scale=0.85,
        transform shape,
        node distance = 2.5cm and 2.5cm,
        every node/.style = {
            draw,
            rectangle,
            align=center,
            font=\small\sffamily
        },
        main/.style = {
            fill=white,
            text width=3.2cm
        },
        auto
    ]

\node[main] (root) {\textbf{MA-LLMs\\For Problem Solving}};

\node[main, below left=1.2cm and 2.0cm of root] (agentProfiles) {\textbf{Agent Profiles}};
\node[main, below right=1.2cm and 2.0cm of root] (decisions) {\textbf{Decision-Making Processes}};

\node[main, below=6.0cm of root] (commStructure) {\textbf{Communication Structure}};

\draw (root) -- (agentProfiles);
\draw (root) -- (decisions);
\draw (root) -- (commStructure);

\node[main, below left=1.89cm and -0.5cm of agentProfiles] (modelFamily) {
  \textbf{Model Family} \\
  \cite{Chen2023ReConcileRC}
};
\node[main, below right=1.89cm and -0.5cm of agentProfiles] (promptEng) {
  \textbf{Prompt-Engineered} \\
  \cite{wei2023chainofthoughtpromptingelicitsreasoning,
  Qian2023CommunicativeAF,
  Wang2023OnTD,
  qianScalingLargeLanguageModelbasedMultiAgent2025,
  xuTowardsReasoningLargeLanguage2023}
};

\draw (agentProfiles) -- (modelFamily);
\draw (agentProfiles) -- (promptEng);

\node[main, below left=1.6cm and 0.1cm of decisions] (majorityVoting) {
  \textbf{Majority Voting} \\
  \cite{Chan2023ChatEvalTB,
  yinExchangeThoughtEnhancingLarge2023,
  Wang2023OnTD,
  liuGroupDebateEnhancingEfficiency2024,
  Chen2023ReConcileRC}
};
\node[main, below=1.6cm of decisions] (consensus) {
  \textbf{Consensus} \\
  \cite{Liang2023EncouragingDT,
  Wang2023OnTD,
  zhang2024exploringcollaborationmechanismsllm,
  Qian2023CommunicativeAF,
  qianScalingLargeLanguageModelbasedMultiAgent2025,
  Xiong2023DivingIT}
};
\node[main, below right=1.6cm and 0.1cm of decisions] (judge) {
  \textbf{Judge} \\
  \cite{Liang2023EncouragingDT,
  Hong2024ArgMedAgentsEC,
  Xiong2023DivingIT,
  wangLargeLanguageModels2023}
};

\draw (decisions) -- (majorityVoting);
\draw (decisions) -- (consensus);
\draw (decisions) -- (judge);

\node[main, below left=1.6cm and 0cm of commStructure]
    (hierNonHier) {
    \textbf{Hierarchical vs.\\Non-Hierarchical} \\
    \cite{xuTowardsReasoningLargeLanguage2023,
    yeMultiAgentSamplingScaling2024}
};
\node[main, below right=1.6cm and 0cm of commStructure]
    (fixVsDyn) {
    \textbf{Fix vs. Dynamic}\\
    \cite{Liang2023EncouragingDT,
    Talebirad2023MultiAgentCH,
    yinExchangeThoughtEnhancingLarge2023}
};
\node[main, below left=1.6cm and 5cm of commStructure]
    (homoVsHetero){
    \textbf{Homogeneous vs.\\Heterogeneous} \\
    \cite{Liang2023EncouragingDT,
    Xiong2023DivingIT,
    Chen2023ReConcileRC}
};
\node[main, below right=1.6cm and 5cm of commStructure]
    (holonic) {
    \textbf{Holonic} \\
    \cite{
    Wang2023OnTD}
};

\draw (commStructure) -- (hierNonHier);
\draw (commStructure) -- (fixVsDyn);
\draw (commStructure) -- (homoVsHetero);
\draw (commStructure) -- (holonic);

\end{tikzpicture}
    \caption{Key characteristics of MA-LLMs for problem solving with citations}
    \Description[Key characteristics of MA-LLMs for problem solving with citations]{In a tree structure the different key characteristics of MA-LLMs for problem-solving are displayed. Each leave is also equipped with the relevant citations. At the root is MA-LLMs for problem-solving. The branch agent profiles contains the leaves model family and prompt-engineered. The branch decision-making processes entails the leaves majority voting, consensus, and judge. The last branch communication structure has four leaves, homogeneous vs. heteogeneous, hierarchical vs non-hierarchical, fix vs. dynamic, and holonic.}
    \label{fig:enter-label}
\end{figure*}

\section{General setup of MA-LLM systems for problem-solving}
\label{sec-general-setup}
Before I answer my research questions later in section~\ref{sec-termination} and section~\ref{sec-scaling}, this section provides an overview of MA-LLMs for problem-solving. Generally, the MA-LLMs provide a single answer to a given task without human assistance in a zero-shot scenario. Each agent is itself an LLM with its intrinsic peculiarities. Some agents might have access to tools such as web browsers, Linux terminals, or Python interpreters. In general, agents exchange information and interact only through text, as opposed to applications that extend beyond problem-solving, such as game engines.
Most approaches to MA-LLMs for problem-solving differ primarily in the \textbf{communication structure} of different agents, the \textbf{decision-making process}, and the \textbf{agent profiles} of individual agents, see figure~\ref{fig:enter-label}. Communication structure broadly refers to the \textit{who-talk-to-whom}, agent profils to the (unique) perspectives of different agents, and the decision-making process to the way of deriving a single answer from the debate among different agents. To describe the specifics regarding communication structure and agent profiles, I primarily rely on the terminology used by \citet{guoLargeLanguageModel2024} and \citet{Multi-Agent-survey} and indicate any changes I have made. Since the agent profiles and the communication structure of MA-LLM systems is not the focus of the work, I do not claim to give a thorough overview of all possible categories of agent profiles and communication structures. I only referenced some papers in the respective classes and not all, which would be beyond the scope of this paper.

\subsection{Agent Profiles}
\label{sec-agentprofiles}

In the introduction, I asked the question: What are valuable perspectives? This question is best answered by the different profiles agents can have. In MA-LLMs, agents are defined by their traits, actions, and skills, which are tailored to meet specific goals. Across various systems, agents assume distinct roles, each with comprehensive descriptions that encompass characteristics, capabilities, behaviors, and constraints write \citet{guoLargeLanguageModel2024}. They also introduce the distinction into three categories of different profile classes, namely: \textbf{pre-defined}, \textbf{model-generated}, and \textbf{data-derived}. In the pre-defined cases, agent profiles are explicitly defined by the system designers. The model-generated method creates agent profiles by models, e.g., large language models. The data-derived method involves constructing agent profiles based on pre-existing datasets. In the special case of problem-solving MA-LLM systems, pre-defined agents are the standard. MA-LLM systems that encompass model-generated agents do exist, yet they are the exception \cite{Talebirad2023MultiAgentCH, chenOptimaOptimizingEffectiveness2025, liuDynamicLLMPoweredAgent2024}. Finally, there exist no data-derived agents in any of the analyzed papers that, e.g., finetuned their agents to have a unique viewpoint or add relevant context to the agent profiles. However, many papers employ different model families in their work, which can be seen as both a form of data-derived profiles and pre-defined agents. As a subcategory of pre-defined agent profiles in MA-LLM systems for problem-solving, I define \textbf{prompt-engineered} agent profiles, the viewpoint and other characteristics are specified in the context window, and \textbf{model families} such as GPT, Claude, and LLAMA among others.

\subsubsection{Prompt-engineering Agents}
\label{sec-meta-prompting}

Prompting techniques, such as Chain-of-Thought, yield strong improvements over baseline performances for single-agent systems and are an active field of research \cite{wei2023chainofthoughtpromptingelicitsreasoning}. Therefore, it is not surprising that similar results can be expected for MA-LLMs. For example, \citet{Qian2023CommunicativeAF} show that removing the profile prompts from all agent's meta-prompts results in the most substantial drop in performance.

However, as \citet{Wang2023OnTD} note, increasing the complexity of the mechanisms may introduce interference between complex [meta-]prompts and complex discussions. \citet{qianScalingLargeLanguageModelbasedMultiAgent2025} distinguish task, profiles, and instructions among others in the analysis of total token consumption.  Arguably, the overall task for which we seek a solution is not part of the agent's profile. However, it is less clear when examining the instructions. Their example is “Add GUI [Graphical User Interface],” which is instead a subtask in software development. Consider examples, such as the one presented by \citet{Wang2023OnTD}, who tell agents to "hold divergent views" or consensus a priori by \citet{xuTowardsReasoningLargeLanguage2023}. I argue that this is also not necessarily part of the agent profile and instead suggests a new category, "instruction," on the same level as agent profiles, topology, and decision-making process with subcategories such as subtask and communication paradigm.

The prompted agent profiles can further be vaguely classified into the incomplete list of the most reoccurring agent that I have found:
\begin{itemize}
    \item Naive agent engaging in a debate, generating ideas 
    \item Judge (Section~\ref{sec-judge})
    \item Red team, criticizer, verifier
    \item Summarizer, orchestrator, secretary, moderator
    \item CEO, CTO, programmer, reviewer, and tester for software development
\end{itemize}

\subsubsection{Model-Families}
\label{sec-model-families}

Employing different model families (GPT, Claude, LLAMA, etc.) or already different training stages of the same model family may seem trivial, but in some papers, it is the primary driver of performance gain \cite{Chen2023ReConcileRC}.  The Section~\ref{sec-homogeneous} on homogeneous vs. heterogeneous agents lists more relevant papers. Typically, the most powerful open-source or API-accessible model families are studied, which I do not list here, as the list would quickly become outdated.

\subsection{Communication Structure (Topology)}
\label{sec-communication-structur}

One of the key design decisions in MA-LLMs systems is the choice of communication structure, or "topology," that governs how agents exchange information and collaborate. Different authors have proposed various taxonomies to characterize these structures, reflecting diverse research directions \cite{yinExchangeThoughtEnhancingLarge2023, guoLargeLanguageModel2024, Multi-Agent-survey}. In this section, I synthesize and compare some of the most prominent classification schemes, highlighting core design principles and examples of their implementation in recent studies. Most rigorously, the topologies of MA-LLM systems are represented using directed graphs, where vertices represent the different agents, and an edge indicates the visibility of an agent's output to other agents. For practicality, I distinguish the four classes \textbf{hierarchical vs. non-hierarchical}, \textbf{fix vs. dynamic}, \textbf{homogeneous vs. heterogeneous}, and \textbf{holonic} topologies following \citet{Multi-Agent-survey}. 

\citet{guoLargeLanguageModel2024} report four classes of graph structure: centralized, decentralized, layered, and shared message pool. \citet{yinExchangeThoughtEnhancingLarge2023} use slightly different graph structures to classify their approaches: bus, star, ring, and tree. Both \citet{guoLargeLanguageModel2024} and \citet{yinExchangeThoughtEnhancingLarge2023} relate the communication structure indirectly or directly to a communication paradigm, which does not transfer further meaning about the functionality of the MA-LLM system. The classification scheme proposed by \citet{Multi-Agent-survey} for multi-agent systems (not specifically multi-agent LLM systems) is both intuitive and comprehensive, allowing for the description of many nuances simultaneously. Although the research community of MA-LLMs has not yet discovered it, the work of \citet{Multi-Agent-survey} warrants more attention. The following Sections~\ref{sec-hierarchy}-\ref{sec-holonic} introduce the part of the classification scheme for multi-agent systems that I find helpful for the description of MA-LLMs.

\subsubsection{Hierarchical vs. Non-Hierarchical}
\label{sec-hierarchy}

A broader lens on multi-agent systems contrasts \textbf{hierarchical} setups with \textbf{non-hierarchical} approaches, which hints at a decision-making mechanism (Section~\ref{sec-termination}). Hierarchical designs grant specific power to some agents, such as a judge \cite{Multi-Agent-survey}. Conversely, non-hierarchical systems distribute equal authority among agents, encouraging each to contribute equally to the solution (Section~\ref{sec-majority-voting}and Section~\ref{sec-consensus}). Some works, such as \citet{xuTowardsReasoningLargeLanguage2023}, empirically explore the effectiveness of flat structures, revealing that while collaboration may emerge naturally, potential deadlocks also arise if the system lacks robust consensus or tie-breaking mechanisms. A special case of hierarchical systems is the tree search by \citet{yeMultiAgentSamplingScaling2024}.

\subsubsection{Fix vs. Dynamic Topologies}
\label{sec-fixed-dynamic}

Another dimension distinguishes \textbf{fix} topologies \cite{Liang2023EncouragingDT}, where the number and connectivity of agents are predetermined, from \textbf{dynamic} topologies, in which the system can spawn or remove agents on demand. Dynamic systems, as discussed by \citet{Talebirad2023MultiAgentCH}, allow for adaptive changes in agent roles and connections based on ongoing performance—agents can even "halt" or deactivate other agents in some proposals. Another example is simply the consistent output termination of agents proposed by \citet{yinExchangeThoughtEnhancingLarge2023}. As soon as an agent gives the same answer twice, they stop receiving or sending information and exit the current communication. Such fluid structures can maximize efficiency and enable MA-LLMs to solve more complex tasks by breaking down a difficult task into easier sub-tasks and letting specialized agents answer them respectively (\textit{schafolding}). Still, they also pose new challenges for ensuring consistency, mainly when the system introduces new agents mid-discussion without a clear handover protocol \cite{Talebirad2023MultiAgentCH}. For more examples of dynamic systems, consult Section~\ref{sec-optimizing-network-structures} on optimizing network structures . 

\subsubsection{Homogeneous vs. Heterogeneous Topologies}
\label{sec-homogeneous}

From another perspective, systems can be labeled \textbf{homogeneous}—using agents with identical capabilities or model architectures—or \textbf{heterogeneous}, where agents differ in size, model family, training data, or agent profiles. Heterogeneous setups can harness the complementary strengths of different foundation models, as exemplified by \citet{Liang2023EncouragingDT, Xiong2023DivingIT}, who organize debates between stronger and weaker LLMs, or \citet{Chen2023ReConcileRC}, who allow different model families to interact with each other to improve solution quality. Despite the added potential of multi-perspective reasoning, heterogeneous systems can have mismatched capabilities. \citet{Xiong2023DivingIT} explain why Comparable LLMs can debate to compromise or adhere to more reasonable perspectives to improve the performance

\subsubsection{Holonic Topologies}
\label{sec-holonic}

\citet{Multi-Agent-survey} propose the category of \textbf{holonic} systems, which are composed of holons of agents grouped according to certain features and communicate with other agents in the same or different holons of the same level. An interesting example of this is the work by \citet{Wang2023OnTD}, where topics are discussed separately in groups and then merged with an additional LLM secretary, allowing for multiple hierarchical levels.

The diversity of possible topologies underscores their importance as a key design consideration in MA-LLM systems. Selecting an appropriate structure affects a system's efficiency and interpretability \cite{reganProblemSolvingLanguageModel2024}. As MA-LLM research matures, developing robust theoretical frameworks for comparing and evaluating different topologies remains an essential open challenge.

\section{Decision-Making Processes}
\label{sec-termination}

In the context of MA-LLM systems for problem-solving, we seek a single solution for the given task. As mentioned in the previous Section~\ref{sec-hierarchy}, hierarchy, and non-hierarchy \ref{sec-hierarchy}, the topology imposes loose constraints on termination or decision-making mechanisms and although social-choice theory, decision theory, and group decision theory can inform decision-making processes for MA-LLMs, the majority of existing approaches employ three primary methods: \textbf{majority voting}, a \textbf{judge}, or \textbf{consensus}. Sometimes, it is unclear whether different papers have the same understanding of consensus vs majority voting. I use consensus throughout the rest of the paper, referring to a situation where no one disagrees with a proposed solution or, even better, when everyone explicitly agrees. Furthermore, almost all papers include a halting mechanism that terminates the process after a fixed number of iterations, thereby avoiding infinite loops.

Note that Arrow's impossibility theorem \cite{Arrow} demonstrates that no universal mechanism can consistently aggregate individual preferences into a single, collective decision while simultaneously satisfying a set of seemingly reasonable criteria. These criteria are foundational to the fairness and rationality of preference aggregation. Arrow's proof demonstrates that any attempt to fulfill all these conditions will inevitably result in a system that fails to meet at least one of them, highlighting the inherent conflict between fairness and decisiveness in collective decision-making.

Viewing the individual agent's solutions to a given problem as their preferences and the solution of the MA-LLMs as the desired aggregate implies that an optimal decision-making process does not exist. To effectively make decisions, loosening one or another restriction gives rise to many different mechanisms that can be better or worse judged by benchmark performances. Unfortunately, I have rarely found papers that compare their MA-LLM systems to those of other MA-LLM systems; instead, they often compare their results only against single-agent LLMs as a baseline. Remember as well that as soon as we leave the narrow definition of problem-solving as I have set it in Section~\ref{sec-problem-solving}, answering which decision-making process is better or worse than another becomes much more difficult because of a lack of a universally accepted metric. 

A notable exception to the three classes, majority voting, judge, and consensus, that I present in more detail is the work by \citet{phamLetModelsSpeak2024}. They empirically find that a good strategy is to use various temperature agents and choose the response of the lower temperature agent as the final debate answer. The paper also stands out in another way. They write that the token sampling step, which is necessary when generating natural language, poses a potential risk of information loss, as it uses only one token to represent the model’s belief across the entire vocabulary. They remove the token sampling step from LLMs and let them communicate their beliefs across the vocabulary through the expectation of the raw transformer output embeddings.

\subsection{Majority Voting}
\label{sec-majority-voting}

Majority voting is one of the most straightforward mechanisms for aggregating multiple outputs in MA-LLM systems. \citet{Chan2023ChatEvalTB} apply a majority-voting scheme to decide the generated responses from different models on open-ended questions and traditional natural language generation tasks, ensuring impartiality and balanced evaluation. Conversely, they average the scores if a task demands a direct numerical assessment. Similarly, \citet{yinExchangeThoughtEnhancingLarge2023} apply majority voting for a homogeneous group of agents. \citet{Wang2023OnTD} propose an interesting constellation in which representatives from nested groups represent the results acquired by majority voting within their respective groups. Secretaries can iteratively deactivate agents that fail to align with the evolving group consensus, underscoring the flexibility of voting-based methods in dynamic topologies. In \citet{liuGroupDebateEnhancingEfficiency2024} upon the conclusion of the debate, all agents reach a consensus, or a majority vote determines the outcome.

There exist more sophisticated voting schemes: \citet{Chen2023ReConcileRC} compare three different voting mechanisms. In their setup, each agent provides its answer along with a confidence estimate. From there, they derive three voting schemes: unweighted majority voting, uncalibrated confi-dence-weighted voting, and calibrated confi-dence-weighted voting, noting that the latter performs slightly better than the former two. Uncalibrated confidence-weighted voting uses the reported confidence score as weights, whereas calibrated confidence weights are the former adjusted to account for the general overconfidence of the LLMs.

\subsection{Judge}
\label{sec-judge}

A second class of decision-making strategies introduces a judge agent to resolve conflicts and guide the outcome. For instance, \citet{Liang2023EncouragingDT} adopt an affirmative vs.\ negative debating structure, where a judge makes the ultimate decision after up to five rounds of argumentation. They also compare it with majority voting to terminate the debate at an optimal juncture to mitigate \textit{Degeneration-Of-Thought} and ensure an efficient resolution. Degeneration-Of-Thought is the phenomenon that once the LLM has established confidence in its solutions, it is unable to generate  novel thoughts later through reflection even if  its initial stance is incorrect. Similarly, \citet{Hong2024ArgMedAgentsEC} propose a constellation of an argument generator and a verifier that iterates multiple times, along with a reasoner equipped with a symbolic solver that makes the final decision. \citet{Xiong2023DivingIT} propose an interesting two-stage decision-making process. For samples that reached a consensus, the conclusion is the consensus stance. When communication terminates after a fixed number of rounds without consensus, they experiment with either majority voting or a judge and conclude that the judge outperforms majority voting. In all these examples, the judicial role concentrates decision-making power in a single entity, providing a clear termination rule that can be highly effective but potentially more susceptible to bias if the judge itself is flawed, e.g., due to positional bias \citet{wangLargeLanguageModels2023}.

\subsection{Consensus}
\label{sec-consensus}

Another approach to decision-making emphasizes achieving consensus. Consensus is a situation where no one disagrees with a proposed solution, or even better when everyone explicitly agrees. \citet{Liang2023EncouragingDT} examine different levels of disagreement by manipulating meta prompts ranging from "must agree" to "must disagree with each other on every point." They find optimal performance emerges at moderately high but not absolute disagreement. By contrast, \citet{Wang2023OnTD} report slightly lower accuracies when prompts explicitly require agents to hold divergent views. They note, however, that the performance drop remains modest.

Building on social psychology principles, \citet{zhang2024exploringcollaborationmechanismsllm} demonstrate that a population composed of "easy-going" (i.e., highly cooperative) agents and "overconfident" ones often achieves consensus more readily. \citet{Qian2023CommunicativeAF} adopt a multi-phase, waterfall-like approach in software development tasks, halting each phase as soon as two agents reach an agreement. Their dual-agent design avoids the complexity of large topologies, thus streamlining negotiation and consensus. This dual-agent setup also enables scaling to multiple rounds of debate and a more significant number of agents. See the Sections~\ref{sec-number-rounds} on rounds of debate and on the number of agents \ref{sec-number-agents}, which \citet{qianScalingLargeLanguageModelbasedMultiAgent2025} analyze in a separate paper.

Nevertheless, consensus-seeking can stall if unanimity proves elusive, a problem reminiscent of grassroots movements where group decisions fail to materialize, thereby underscoring how social-choice theory and related fields may inspire more robust and flexible strategies, such as two-stage decision-making process proposed by \citet{Xiong2023DivingIT}, for future MA-LLM systems.

\section{Scaling}
\label{sec-scaling}

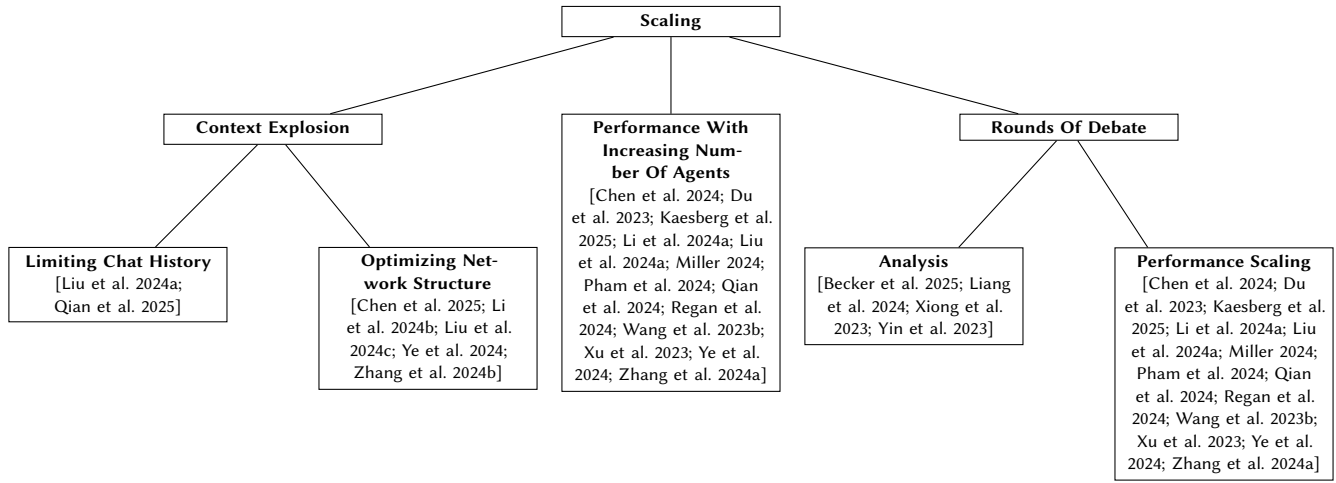
\begin{figure*}
    \hspace{0cm}
\begin{tikzpicture}[
        xshift=-2cm,
        scale=0.85,
        transform shape,
        node distance = 2.5cm and 2.5cm,
        every node/.style = {
            draw,
            rectangle,
            align=center,
            font=\small\sffamily
        },
        main/.style = {
            fill=white,
            text width=3.2cm
        },
        auto
    ]

\node[main] (root) {\textbf{Scaling}};

\node[main, below left=1.2cm and 2.8cm of root] (context) {\textbf{Context Explosion}};
\node[main, below right=1.2cm and 2.8cm of root] (rounds) {\textbf{Rounds Of Debate}};
\node[main, below=1.2cm of root] (performance) {\textbf{Performance With Increasing Number Of Agents}   \\\cite{millerAddingErrorBars2024, xuTowardsReasoningLargeLanguage2023, liuGroupDebateEnhancingEfficiency2024, liImprovingMultiAgentDebate2024, zhang2024exploringcollaborationmechanismsllm, kaesbergVotingConsensusDecisionMaking2025, reganProblemSolvingLanguageModel2024, Wang2023OnTD, Chen2023ReConcileRC, duImprovingFactualityReasoning2023, yeMultiAgentSamplingScaling2024, Qian2023CommunicativeAF, phamLetModelsSpeak2024}};

\draw (root) -- (context);
\draw (root) -- (rounds);
\draw (root) -- (performance);

\node[main, below left=1.6cm and -1.0cm of context] (limiting) {
  \textbf{Limiting Chat History} \\
  \cite{liuGroupDebateEnhancingEfficiency2024, qianScalingLargeLanguageModelbasedMultiAgent2025}
};
\node[main, below right=1.6cm and -1.0cm of context] (optimizing) {
  \textbf{Optimizing Network Structure} \\
  \cite{liuDynamicLLMPoweredAgent2024, liSMoAImprovingMultiagent2024, zhangCutCrapEconomical2024, chenOptimaOptimizingEffectiveness2025, yeMultiAgentSamplingScaling2024}
};

\draw (context) -- (limiting);
\draw (context) -- (optimizing);

\node[main, below left=1.67cm and -1.0cm of rounds] (analysis) {
  \textbf{Analysis} \\
  \cite{beckerStayFocusedProblem2025, Liang2023EncouragingDT, Xiong2023DivingIT, yinExchangeThoughtEnhancingLarge2023}
};
\node[main, below right=1.67cm and -1.0cm of rounds] (scaling) {
  \textbf{Performance Scaling} \\
  \cite{millerAddingErrorBars2024, xuTowardsReasoningLargeLanguage2023, liuGroupDebateEnhancingEfficiency2024, liImprovingMultiAgentDebate2024, zhang2024exploringcollaborationmechanismsllm, kaesbergVotingConsensusDecisionMaking2025, reganProblemSolvingLanguageModel2024, Wang2023OnTD, Chen2023ReConcileRC, duImprovingFactualityReasoning2023, yeMultiAgentSamplingScaling2024, Qian2023CommunicativeAF, phamLetModelsSpeak2024}
};

\draw (rounds) -- (analysis);
\draw (rounds) -- (scaling);

\end{tikzpicture}
    \caption{Scaling in MA-LLMs with Citations}
    \Description[Scaling in MA-LLMs with Citations]{The figure displays a tree structure contating the different aspects in which I classify the scaling of MY-LLMs for problem-solving and relevant citations. At the root is the node scaling. Then there are three edges to the nodes context eplosion, performance with increasing number of agents, and rounds of debate. Context explosiong has the two nodes limiting chat history and optimizing network structure, and the node rounds of debate has the two leaves analysis, and performance scaling.}
    \label{fig:graph}
\end{figure*}

Scaling in MA-LLMs concerns the challenges and opportunities associated with increasing the number of participating agents and the communication rounds within agent interactions. This section first addresses the implications of increased \textbf{communication volume}, often referred to as \textit{context explosion}, and then discusses how performance metrics of MA-LLM systems vary with the \textbf{number of agents} and \textbf{debate rounds}, see figure~\ref{fig:graph}.

\subsection{Communication Volume}
\label{sec-communication-volume}

The scaling of communication volume, measured as the token count per agent input, can critically limit the scalability of MA-LLM systems. Increased communication complexity has been identified as a potential bottleneck, leading to diminishing marginal returns as the system approaches the functional limits of individual LLM capabilities \citep{Wang2023OnTD}. Complex prompts and extensive agent interactions contribute to this phenomenon by increasing context overhead, potentially causing interference rather than improvements.

The term \textit{context explosion}, first introduced by \citet{liuAutonomousAgentsCollaborative2024}, describe explicitly the growth in context required by specific multi-agent configurations. In a rigorous mathematical treatment, \citet{qianScalingLargeLanguageModelbasedMultiAgent2025} provide a formal derivation for context explosion under particular conditions, such as the fully connected agent topologies with a judge at the end who receives the entire chat history of all agents. They demonstrate that context pressure scales quadratically with the number of agents for the judge and further depends on variables like task length, profile length, instruction length, average output length, and the maximum number of debate rounds. Complementing these insights, \citet{yinExchangeThoughtEnhancingLarge2023} empirically compare communication volume scaling across different agent constellations, highlighting significant variation.

To mitigate the challenges posed by context explosion, I categorize current strategies broadly into two methods: \textbf{optimizing chat history} and \textbf{optimizing network structure}.

\subsubsection{Limiting Chat History}
\label{sec-limiting-chat-history}

Optimizing chat history involves techniques designed to reduce redundant communication among agents. \citet{liuGroupDebateEnhancingEfficiency2024} propose \textit{GroupDebate}, a structured method of dividing agents into subgroups that internally debate and produce summaries. These condensed summaries are then shared across groups, significantly reducing the token cost per communication round. Additionally, \citet{qianScalingLargeLanguageModelbasedMultiAgent2025} integrate short-term and long-term memory management techniques to control context visibility efficiently, enabling agents to reference past interactions without extensive token overhead.

\subsubsection{Optimizing Network Structure}
\label{sec-optimizing-network-structures}

Network structure optimization strategies aim to enhance communication efficiency by dynamically managing the interactions between agents. \citet{liuDynamicLLMPoweredAgent2024} introduce the \textit{Dynamic LLM-Powered Agent Network} (DyLAN), a two-phase process of team optimization and task solving. Initially, DyLAN employs an unsupervised metric, the Agent Importance Score, to select a subset of the most effective agents from a larger pool. Subsequently, these agents collaborate dynamically, with lower-performing agents being selectively deactivated in real time by an LLM-driven ranking mechanism.  Further developments include \textit{Sparse Mixture-of-Agents} by \citet{liSMoAImprovingMultiagent2024}, which also employs LLM-driven response selection and early stopping. Similarly, \citet{zhangCutCrapEconomical2024} propose \textit{AgentPrune}, a method explicitly targeting redundancy in agent communication by performing efficient one-shot pruning on the communication graph. \citet{chenOptimaOptimizingEffectiveness2025} demonstrate that optimized agent configurations can be learned directly from data, albeit with limited transferability across tasks.

Tree-based methods represent another innovative approach to context management. \citet{yeMultiAgentSamplingScaling2024} conceptualize multi-agent interactions as dynamic multi-step decision-making processes, utilizing Monte Carlo Tree Search (MCTS) to prune less promising response paths based on reward heuristics selectively. This method optimizes the generation structure dynamically, significantly reducing context requirements without sacrificing response quality.

A comprehensive examination of these structural variations and their implications is provided in the dedicated section~\ref{sec-communication-structur} on topology.

\subsection{Number of Debate Rounds}
\label{sec-number-rounds}

The number of debate rounds in MA-LLM systems significantly impacts problem-solving performance, exhibiting an intricate balance between facilitating adequate information exchange and avoiding cognitive overload or drift. I divide this section into analyzing the emergent behavior that manifests itself through prolonged debates and the performance scaling.

\subsubsection{Analysis}
\citet{beckerStayFocusedProblem2025} introduce the concept of "problem drift," referring to the systematic decay in performance as agents progressively diverge from the original task across multiple debate rounds. They report that approximately 0.5\% of analyzed discussions benefit from the multi-agent debate and improve performance compared to the first turn’s draft, while approximately 0.8\% suffers from a performance drop. This phenomenon highlights a critical limitation of extended interactions, as the benefit of additional exchanges can diminish or reverse entirely.

The optimal number of debate rounds varies depending on task complexity. \citet{Liang2023EncouragingDT} observe that complex tasks benefit from multiple rounds of debate, as nuanced information accumulates gradually, whereas more straightforward tasks typically reach peak performance after just one or two rounds. They propose an "adaptive break" strategy, indicating that terminating the debate at an optimal moment can maximize accuracy and mitigate the risk of prolonged or repetitive discussions. Similarly, \citet{Xiong2023DivingIT} track inter-agent inconsistencies over multiple rounds and find that agreements among agents typically solidify within two to three rounds, especially among similarly capable LLMs. Conversely, more challenging scenarios occasionally necessitate extended debate. \citet{yinExchangeThoughtEnhancingLarge2023} show that most cases reach consensus within three rounds but allow agents to engage in more rounds of debate on questions where consensus is challenging to achieve.

\subsubsection{Performance Scaling}
Increasing rounds of debate can initially enhance performance by enabling more profound exchanges of perspectives and feedback, but too many rounds introduce risks of diminishing returns or even performance degradation. Generally, many studies lack reports of standard errors in their benchmark performances, complicating statistical analysis \citep{millerAddingErrorBars2024}. Furthermore, many studies only report benchmark performances up to three or four rounds of debate, limiting the analysis of performance trends across a more significant number of rounds.

Extended debate rounds can negatively impact performance, as observed in several empirical studies. \citet{xuTowardsReasoningLargeLanguage2023} note an improvement in accuracy only until the second round, after which accuracy declines. Similarly, \citet{liuGroupDebateEnhancingEfficiency2024} and \citet{liImprovingMultiAgentDebate2024} report a peak in accuracy around four rounds, followed by a slight decline.

Moreover, fluctuations in performance highlight a nuanced relationship between debate rounds and accuracy. \citet{zhang2024exploringcollaborationmechanismsllm} illustrate that initial rounds substantially improve performance, but further rounds introduce volatility, particularly between three and ten rounds. \citet{kaesbergVotingConsensusDecisionMaking2025} also observe a slight downward trend in performance with increasing debate rounds, accompanied by an abrupt improvement from nine to ten rounds. \citet{reganProblemSolvingLanguageModel2024} compare network topologies, such as fully connected and disconnected networks, highlighting how information flow impacts accuracy.

\citet{Wang2023OnTD} explore one to three rounds and report a steep accuracy gain when agents' meta-prompts include "hold different views" before converging on a final response, suggesting controlled disagreement can refine complex arguments. Similarly, \citet{Chen2023ReConcileRC} show performance improvements saturate around three rounds. \citet{duImprovingFactualityReasoning2023} corroborate these findings, reporting accuracy plateaus around four rounds, and \citet{yeMultiAgentSamplingScaling2024} demonstrate performance saturation after five rounds using Monte Carlo Tree Search (MCTS).

\citet{Qian2023CommunicativeAF} find that a waterfall model structure consistently improves outcomes in software development, indicating topology and task-specific considerations are crucial in selecting debate lengths. \citet{phamLetModelsSpeak2024} also report that performance improves until three debate rounds, after which it fluctuates slightly at a high level.

Collectively, these studies underscore the complexity of optimizing debate rounds in MA-LLM systems. Increased interactions enhance initial performance, yet careful management is necessary to prevent diminishing returns or problem drift. Optimal debate rounds depend heavily on task characteristics, network topology, and specific meta-prompting strategies, highlighting this as a critical area for future research.

\subsection{Number of Agents}
\label{sec-number-agents}

This subsection synthesizes research findings on how the performance of MA-LLMs systems scales with an increasing number of participating agents. Although adding more agents can introduce diverse perspectives, this also raises the risk of redundancy and contradictions. Notably, many studies in this area do not consistently report standard errors or other measures of statistical significance, complicating precise assessments of scalability trends.

Several studies investigate whether an optimal team size exists for maximizing performance. \citet{Liang2023EncouragingDT} observe only modest performance degradation when the number of agents grows from two to four, attributing this decline to the LLM’s constrained capacity for concurrently tracking multiple viewpoints. \citet{juFloodingSpreadManipulated2024} reinforce this caution, reporting a significant drop in accuracy when scaling from five to ten agents. For tasks such as text summarization, four-agent constellations have been reported to outperform both smaller and larger configurations\citet{Wu2023LargeLM}. \citet{Wu2023LargeLM} also argue that fewer agents may lack sufficient diversity of viewpoints, while larger teams dilute individual contributions, diminishing the overall focus. \citet{Wu2023LargeLM}  and \citet{xuTowardsReasoningLargeLanguage2023} find that performance peaks at four agents.

\citet{Chan2023ChatEvalTB} identify three to four debaters as the optimal configuration, balancing effective cross-verification and manageable complexity. \citet{zhang2024exploringcollaborationmechanismsllm} corroborate this, identifying three agents as an ideal number, optimizing the balance between decision quality, consensus ease, and computational demands. \citet{yeMultiAgentSamplingScaling2024} demonstrate, via Monte Carlo Tree Search (MCTS), that additional diversity in model types beyond two to three agents fails to yield further performance gains. 

In contrast, \citet{duImprovingFactualityReasoning2023} find continuous improvement in mathematical accuracy with up to seven agents, suggesting potential task-specific scalability. \citet{liuGroupDebateEnhancingEfficiency2024} extend this observation up to eight agents, reporting diminishing accuracy improvements without clear evidence of plateauing. \citet{zhangGDesignerArchitectingMultiagent2025}, experimenting with agent groups of 5, 10, and 20, document constant incremental gains, and propose adaptive solutions to optimize agent communication and efficiency through dynamic topologies. Further evidence comes from \citet{zhangCutCrapEconomical2024}, who observe significant improvements in performance between three to five agents but only marginal gains from further scaling up to nine agents. 

\citet{zhang2024exploringcollaborationmechanismsllm} highlight the unique and obvious advantage of using an odd number of participants to avoid recurring ties. They also find that the performance differences among groups of odd numbers are minor, thereby recommending three agents as a sweet spot balancing quality, consensus, and computational cost.

Finally, \citet{liMoreAgentsAll2024} present one of the first systematic studies of raw LLM agent scaling, observing steady but modest performance increases up to 40 agents, primarily through simple sampling and voting mechanisms without debate. In a similar vein, \citet{qianScalingLargeLanguageModelbasedMultiAgent2025} introduce MACNET, a scalable framework that supports collaboration among over a thousand agents, revealing a logarithmic growth in performance with irregular agent topologies surpassing regular ones. They achieve that through excessively capping the forwarded chat history.

\citet{Talebirad2023MultiAgentCH} highlight that autonomously spawning agents can amplify conflicts without proper management, suggesting a non-trivial ceiling on the benefits of unbounded growth. Additionally, some system configurations inherently do not scale at all, such as the circular structure of \citep{Hong2024ArgMedAgentsEC} with a generator, verifier, and reasoner, all of which have specific instructions and only communicate with the next one.

Collectively, these findings emphasize that while scaling agent numbers generally enhances robustness and accuracy, practical limitations and diminishing returns frequently emerge. Optimal agent count varies significantly with task complexity, agent diversity, and communication topology, necessitating adaptive management for effective scalability.

\section{Epilogue}
\subsection{Discussion}
\label{sec-discussion}

MA-LLMs provide notable advantages over single-agent LLMs by integrating diverse perspectives and iterative feedback, enhancing problem-solving effectiveness. Nevertheless, MA-LLMs face challenges such as increased computational costs and complexity in coordination and decision-making. This section first discusses general aspects of MA-LLM implementation, followed by specific analyses: RQ1 on decision-making processes (majority voting, judges, consensus) and RQ2 on performance scaling relative to agent numbers and interaction rounds.

Emergent behaviors and scaling dynamics may constitute a functional post-hoc classification in MA-LLMs, supplementing apriori categories like agent profiles or topology. \citet{guoLargeLanguageModel2024} emphasize communication paradigms as a fundamental categorization dimension, idealizing truth-seeking as a shared objective. However, analogous to human interactions, communication in MA-LLMs is susceptible to individual and group biases that arise intrinsically from the architecture, topology, or individual agent characteristics yet only manifest themself during and after the debate, further illustrating the concept of post-hoc classification analysis. Personal biases, such as positional bias found specifically in MA-LLMs for problem-solving and more broadly recognized biases like racial or gender biases prevalent in LLMs, remain an active area of research. Interestingly, while individual biases are extensively explored, group-level biases, with the exceptions of Degeneration-of-Thought and Problem-Drift, are comparatively under-researched.

In the context of agent profiles, data-derived agent profiles remain notably absent from the current academic literature, although they are likely present in real-world implementations. Almost universally, current MA-LLM systems do not adopt holistic training strategies for either the entire multi-agent system or individual agents, primarily due to prohibitive training costs and complexity. An exception is the Optima system developed by \citet{chenOptimaOptimizingEffectiveness2025}, highlighting a rare example of such comprehensive training. Additionally, reinforcement learning methodologies have been notably excluded from the scope of this review, suggesting another avenue for future exploration.

Current research raises the question of whether predefined agent profiles sufficiently capture all valuable perspectives for complex problem-solving tasks. It appears to that perspectives are tailored primarily in fields such as software development and text summarization, with limited customization observed in other problem-solving contexts.

The utility of classification schemes from traditional multi-agent systems literature, such as \citet{Multi-Agent-survey}, for the MA-LLM research community remains to be seen. By bridging this gap, my work can benefit future research that more profoundly explores the adaptation of these classification schemes to better address the specific challenges and opportunities inherent in MA-LLM systems.

\paragraph{\textbf{Research Questions 1:}}
\begin{itemize}
    \item How do MA-LLMs terminate discussions and make decisions? What are the strengths and limitations of established methods
\end{itemize}

Determining an optimal decision-making process or terminology is highly challenging due to the immense variety of available methods. While practical applications may benefit from adopting decision-making processes drawn directly from real-world contexts, novel approaches such as Monte Carlo decision trees \cite{yeMultiAgentSamplingScaling2024} and the lowest temperature principle \cite{phamLetModelsSpeak2024} illustrate innovative cross-disciplinary or unique MA-LLM-specific solutions. Nevertheless, consensus and judge-based decision-making approaches, although straightforward, provide limited flexibility for adaptation. Voting mechanisms, by contrast, present a vast array of possibilities, ranging from weighted voting schemes, as demonstrated by \citet{Chen2023ReConcileRC}, to ranked-choice or approval voting methods, each offering distinct advantages and challenges.

\paragraph{\textbf{Research Questions 2:}}
\begin{itemize}
    \item How do multi-agent LLM systems scale with the number of agents and rounds of debate with respect to their problem-solving performance? What are the challenges arising from scaling MA-LLMs?
\end{itemize}

In theory, increasing the number of agents or rounds of debate is expected to either steadily enhance performance or reach a plateau with diminishing returns. However, empirical studies indicate a more complex relationship. Initially, adding agents or debate rounds consistently boosts performance; however, both factors can eventually lead to performance degradation due to phenomena such as problem drift, interference, or redundancy. The exact point at which scaling becomes detrimental varies significantly based on topology, task complexity, and system configurations. Open-ended tasks such as software development and text summarization benefit more from increased agent numbers compared to more straightforward tasks like question-answering, indicating task-specific sensitivity. Overall, while evidence suggests that additional agents rarely lead to significant performance drops and often result in relatively marginal improvement or saturation, however increasing debate rounds beyond an optimal point usually leads to declining performance.

Rigorous statistical analysis of benchmark performance in MA-LLM systems remains notably weak. Most studies fail to report standard errors or critically examine the assumption of independent and identically distributed tasks, making robust meta-analytical comparisons challenging. Conducting extensive empirical evaluations of MA-LLM systems is resource-intensive and may necessitate the development of novel validation methodologies to generalize findings effectively.

Inference computational cost presents another substantial challenge in the research of MA-LLM systems. As the number of agents and debate rounds increases, token usage escalates sharply, directly impacting computational resources and associated costs. While inference costs have reduced over time \cite{epoch2025llminferencepricetrends}, they continue to pose significant limitations for comprehensive academic investigations. \citet{liuGroupDebateEnhancingEfficiency2024} specifically identify token cost growth as a critical obstacle, underscoring the need to develop efficient communication strategies to sustain MA-LLM scalability and performance improvements.

\subsection{Conclusion}
\label{sec-conclusion}

This review explores the growing research field of multi-agent large language models (MA-LLMs), showing how multiple interacting agents often perform better than single-agent systems across a range of problem-solving tasks. By combining together insights from traditional multi-agent frameworks and recent research on LLMs, the review emphasizes the significant impact of agent profiles, communication structures, and decision-making processes on overall performance. While predefined roles for agents are currently common practice, there is potential in developing specialized or data-driven agent roles.

A central challenge identified is the balance between scaling up the number of agents or interactions to enhance performance and managing the associated computational costs and potential difficulties that arise from it. Recent work exploring dynamic network configurations and pruning strategies presents promising methods for addressing these issues. The literature also highlights bias mitigation as a critical concern, noting that biases can emerge at both the individual and group levels, adversely affecting results.

Although the field has made impressive progress, a strong need remains for rigorous statistical analyses and more direct comparisons between different MA-LLM systems. Filling these gaps can provide clearer guidelines on how best to balance performance gains against computational costs. Overall, the review highlights the considerable potential of MA-LLM approaches for complex tasks, while emphasizing that continued methodological innovation and thorough evaluation are crucial.

\subsection{Future Work}
\label{sec-limitations}

A pressing need in MA-LLM research is a more thorough investigation of cost-effectiveness, as both agent count and debate rounds significantly influence token usage and, consequently, computational expense. Studies by \citet{liuGroupDebateEnhancingEfficiency2024} and \citet{zhang2024exploringcollaborationmechanismsllm} demonstrate the feasibility of quantifying this trade-off, but a more formal, comparative framework is required. Equally crucial is the design of rigorous meta-analyses comparing various MA-LLM architectures and decision mechanisms; such research would bridge the current gap where MA-LLMs are predominantly benchmarked only against single-agent baselines. Another promising direction is the development of methods for large-scale collaboration in complex, dynamic environments, potentially orchestrating hundreds or even thousands of agents to tackle real-world tasks such as extensive software engineering or intricate multi-step problem-solving. Finally, advancements in evaluation methods—especially those capturing emergent behaviors, social biases, and agent synergy—would provide deeper insights into how different MA-LLM configurations can be optimized for both effectiveness and fairness. By addressing these four focal areas, the research community can establish a more cohesive understanding of MA-LLM systems, paving the way for robust, scalable, and cost-efficient multi-agent solutions.

\begin{acks}
To Jonas Becker for supervision and support. 
\end{acks}

\bibliographystyle{ACM-Reference-Format}
\bibliography{sample-sigconf}


\begin{thebibliography}{49}


\ifx \showCODEN    \undefined \def \showCODEN     #1{\unskip}     \fi
\ifx \showDOI      \undefined \def \showDOI       #1{#1}\fi
\ifx \showISBNx    \undefined \def \showISBNx     #1{\unskip}     \fi
\ifx \showISBNxiii \undefined \def \showISBNxiii  #1{\unskip}     \fi
\ifx \showISSN     \undefined \def \showISSN      #1{\unskip}     \fi
\ifx \showLCCN     \undefined \def \showLCCN      #1{\unskip}     \fi
\ifx \shownote     \undefined \def \shownote      #1{#1}          \fi
\ifx \showarticletitle \undefined \def \showarticletitle #1{#1}   \fi
\ifx \showURL      \undefined \def \showURL       {\relax}        \fi
\providecommand\bibfield[2]{#2}
\providecommand\bibinfo[2]{#2}
\providecommand\natexlab[1]{#1}
\providecommand\showeprint[2][]{arXiv:#2}

\bibitem[Arrow(1950)]%
        {Arrow}
\bibfield{author}{\bibinfo{person}{Kenneth~J. Arrow}.}
  \bibinfo{year}{1950}\natexlab{}.
\newblock \showarticletitle{{A Difficulty in the Concept of Social Welfare}}.
\newblock \bibinfo{journal}{\emph{Journal of Political Economy}}
  \bibinfo{volume}{58}, \bibinfo{number}{4} (\bibinfo{year}{1950}),
  \bibinfo{pages}{328--328}.
\newblock
\urldef\tempurl%
\url{https://doi.org/10.1086/256963}
\showDOI{\tempurl}


\bibitem[Aschenbrenner(2024)]%
        {IntroductionSITUATIONALAWARENESS}
\bibfield{author}{\bibinfo{person}{Leopold Aschenbrenner}.}
  \bibinfo{year}{2024}\natexlab{}.
\newblock \bibinfo{title}{Introduction - {{SITUATIONAL AWARENESS}}: {{The
  Decade Ahead}}}.
\newblock \bibinfo{howpublished}{\url{https://situational-awareness.ai/}}.
\newblock


\bibitem[Becker et~al\mbox{.}(2025)]%
        {beckerStayFocusedProblem2025}
\bibfield{author}{\bibinfo{person}{Jonas Becker},
  \bibinfo{person}{Lars~Benedikt Kaesberg}, \bibinfo{person}{Andreas Stephan},
  \bibinfo{person}{Jan~Philip Wahle}, \bibinfo{person}{Terry Ruas}, {and}
  \bibinfo{person}{Bela Gipp}.} \bibinfo{year}{2025}\natexlab{}.
\newblock \bibinfo{title}{Stay Focused: Problem Drift in Multi-Agent Debate}.
\newblock
\newblock
\showeprint[arxiv]{2502.19559}~[cs.CL]
\urldef\tempurl%
\url{https://arxiv.org/abs/2502.19559}
\showURL{%
\tempurl}


\bibitem[Brown et~al\mbox{.}(2024)]%
        {meta-benchmark}
\bibfield{author}{\bibinfo{person}{Sam Brown}, \bibinfo{person}{Basil Labib},
  \bibinfo{person}{Codruta Lugoj}, {and} \bibinfo{person}{Sai Sasank~Y}.}
  \bibinfo{year}{2024}\natexlab{}.
\newblock \bibinfo{title}{{A}uto-{E}nhance: {D}eveloping a meta-benchmark to
  measure {L}{L}{M} agents’ ability to improve other agents — {A}{I}
  {A}lignment {F}orum --- alignmentforum.org}.
\newblock
  \bibinfo{howpublished}{\url{https://www.alignmentforum.org/posts/s9zd6f9eZ8qN2jrcu/auto-enhance-developing-a-meta-benchmark-to-measure-llm}}.
\newblock
\newblock
\shownote{[Accessed 25-11-2024]}.


\bibitem[Chan et~al\mbox{.}(2023)]%
        {Chan2023ChatEvalTB}
\bibfield{author}{\bibinfo{person}{Chi-Min Chan}, \bibinfo{person}{Weize Chen},
  \bibinfo{person}{Yusheng Su}, \bibinfo{person}{Jianxuan Yu},
  \bibinfo{person}{Wei Xue}, \bibinfo{person}{Shanghang Zhang},
  \bibinfo{person}{Jie Fu}, {and} \bibinfo{person}{Zhiyuan Liu}.}
  \bibinfo{year}{2023}\natexlab{}.
\newblock \bibinfo{title}{ChatEval: Towards Better LLM-based Evaluators through
  Multi-Agent Debate}.
\newblock
\newblock
\showeprint[arxiv]{2308.07201}~[cs.CL]
\urldef\tempurl%
\url{https://arxiv.org/abs/2308.07201}
\showURL{%
\tempurl}


\bibitem[Chen et~al\mbox{.}(2024)]%
        {Chen2023ReConcileRC}
\bibfield{author}{\bibinfo{person}{Justin~Chih{-}Yao Chen},
  \bibinfo{person}{Swarnadeep Saha}, {and} \bibinfo{person}{Mohit Bansal}.}
  \bibinfo{year}{2024}\natexlab{}.
\newblock \showarticletitle{{ReConcile: Round-Table Conference Improves
  Reasoning via Consensus among Diverse LLMs}}. In
  \bibinfo{booktitle}{\emph{{Proceedings of the 62nd Annual Meeting of the ACL
  (Volume 1: Long Papers)}}}. \bibinfo{publisher}{Association for Computational
  Linguistics}, \bibinfo{address}{Online}, \bibinfo{pages}{7066--7085}.
\newblock


\bibitem[Chen et~al\mbox{.}(2025)]%
        {chenOptimaOptimizingEffectiveness2025}
\bibfield{author}{\bibinfo{person}{Weize Chen}, \bibinfo{person}{Jiarui Yuan},
  \bibinfo{person}{Chen Qian}, \bibinfo{person}{Cheng Yang},
  \bibinfo{person}{Zhiyuan Liu}, {and} \bibinfo{person}{Maosong Sun}.}
  \bibinfo{year}{2025}\natexlab{}.
\newblock \bibinfo{booktitle}{\emph{Optima: Optimizing Effectiveness and
  Efficiency for LLM-Based Multi-Agent System}}.
\newblock Tsinghua University.
\newblock
\urldef\tempurl%
\url{https://doi.org/10.48550/arXiv.2410.08115}
\showDOI{\tempurl}
\showeprint[arXiv]{2410.08115}~[cs]


\bibitem[Christiano et~al\mbox{.}(2023)]%
        {christiano2023deepreinforcementlearninghuman}
\bibfield{author}{\bibinfo{person}{Paul Christiano}, \bibinfo{person}{Jan
  Leike}, \bibinfo{person}{Tom~B. Brown}, \bibinfo{person}{Miljan Martic},
  \bibinfo{person}{Shane Legg}, {and} \bibinfo{person}{Dario Amodei}.}
  \bibinfo{year}{2023}\natexlab{}.
\newblock \bibinfo{title}{Deep reinforcement learning from human preferences}.
\newblock
\newblock
\showeprint[arxiv]{1706.03741}~[stat.ML]
\urldef\tempurl%
\url{https://arxiv.org/abs/1706.03741}
\showURL{%
\tempurl}


\bibitem[Cottier et~al\mbox{.}(2025)]%
        {epoch2025llminferencepricetrends}
\bibfield{author}{\bibinfo{person}{Ben Cottier}, \bibinfo{person}{Ben Snodin},
  \bibinfo{person}{David Owen}, {and} \bibinfo{person}{Tom Adamczewski}.}
  \bibinfo{year}{2025}\natexlab{}.
\newblock \bibinfo{title}{LLM inference prices have fallen rapidly but
  unequally across tasks}.
\newblock
\newblock
\urldef\tempurl%
\url{https://epoch.ai/data-insights/llm-inference-price-trends}
\showURL{%
\tempurl}
\newblock
\shownote{Accessed: 2025-03-28}.


\bibitem[Du et~al\mbox{.}(2023)]%
        {duImprovingFactualityReasoning2023}
\bibfield{author}{\bibinfo{person}{Yilun Du}, \bibinfo{person}{Shuang Li},
  \bibinfo{person}{Antonio Torralba}, \bibinfo{person}{Joshua~B. Tenenbaum},
  {and} \bibinfo{person}{Igor Mordatch}.} \bibinfo{year}{2023}\natexlab{}.
\newblock \bibinfo{title}{Improving {{Factuality}} and {{Reasoning}} in
  {{Language Models}} through {{Multiagent Debate}}}.
\newblock
\newblock
\urldef\tempurl%
\url{https://doi.org/10.48550/arXiv.2305.14325}
\showDOI{\tempurl}
\showeprint[arxiv]{2305.14325}


\bibitem[{Elicit}(2023)]%
        {elicit}
\bibfield{author}{\bibinfo{person}{{Elicit}}.} \bibinfo{year}{2023}\natexlab{}.
\newblock \bibinfo{booktitle}{\emph{Elicit: The AI Research Assistant}}.
\newblock Ought.org.
\newblock
\urldef\tempurl%
\url{https://elicit.com}
\showURL{%
\tempurl}


\bibitem[{Epoch AI}(2024)]%
        {EpochNotableModels2024}
\bibfield{author}{\bibinfo{person}{{Epoch AI}}.}
  \bibinfo{year}{2024}\natexlab{}.
\newblock \bibinfo{title}{Data on Notable AI Models}.
\newblock
\newblock
\urldef\tempurl%
\url{https://epoch.ai/data/notable-ai-models}
\showURL{%
\tempurl}
\newblock
\shownote{Accessed: 2024-11-25}.


\bibitem[Erdil and Besiroglu(2023)]%
        {erdilAlgorithmicProgressComputer2023}
\bibfield{author}{\bibinfo{person}{Ege Erdil} {and} \bibinfo{person}{Tamay
  Besiroglu}.} \bibinfo{year}{2023}\natexlab{}.
\newblock \bibinfo{title}{Algorithmic Progress in Computer Vision}.
\newblock
\newblock
\urldef\tempurl%
\url{https://doi.org/10.48550/arXiv.2212.05153}
\showDOI{\tempurl}
\showeprint[arxiv]{2212.05153}


\bibitem[Glazer et~al\mbox{.}(2024)]%
        {glazerFrontierMathBenchmarkEvaluating2024}
\bibfield{author}{\bibinfo{person}{Elliot Glazer}, \bibinfo{person}{Ege Erdil},
  \bibinfo{person}{Tamay Besiroglu}, \bibinfo{person}{Diego Chicharro},
  \bibinfo{person}{Evan Chen}, \bibinfo{person}{Alex Gunning},
  \bibinfo{person}{Caroline~Falkman Olsson}, \bibinfo{person}{Jean-Stanislas
  Denain}, \bibinfo{person}{Anson Ho}, \bibinfo{person}{Emily de~Oliveira
  Santos}, \bibinfo{person}{Olli J{\"a}rviniemi}, \bibinfo{person}{Matthew
  Barnett}, \bibinfo{person}{Robert Sandler}, \bibinfo{person}{Matej Vrzala},
  \bibinfo{person}{Jaime Sevilla}, \bibinfo{person}{Qiuyu Ren},
  \bibinfo{person}{Elizabeth Pratt}, \bibinfo{person}{Lionel Levine},
  \bibinfo{person}{Grant Barkley}, \bibinfo{person}{Natalie Stewart},
  \bibinfo{person}{Bogdan Grechuk}, \bibinfo{person}{Tetiana Grechuk},
  \bibinfo{person}{Shreepranav~Varma Enugandla}, {and} \bibinfo{person}{Mark
  Wildon}.} \bibinfo{year}{2024}\natexlab{}.
\newblock \bibinfo{title}{{{FrontierMath}}: {{A Benchmark}} for {{Evaluating
  Advanced Mathematical Reasoning}} in {{AI}}}.
\newblock
\newblock
\urldef\tempurl%
\url{https://doi.org/10.48550/arXiv.2411.04872}
\showDOI{\tempurl}
\showeprint[arxiv]{2411.04872}


\bibitem[Guo et~al\mbox{.}(2024)]%
        {guoLargeLanguageModel2024}
\bibfield{author}{\bibinfo{person}{Taicheng Guo}, \bibinfo{person}{Xiuying
  Chen}, \bibinfo{person}{Yaqi Wang}, \bibinfo{person}{Ruidi Chang},
  \bibinfo{person}{Shichao Pei}, \bibinfo{person}{Nitesh~V. Chawla},
  \bibinfo{person}{Olaf Wiest}, {and} \bibinfo{person}{Xiangliang Zhang}.}
  \bibinfo{year}{2024}\natexlab{}.
\newblock \showarticletitle{Large Language Model Based Multi-agents: A Survey
  of Progress and Challenges}. In \bibinfo{booktitle}{\emph{Proceedings of the
  Thirty-Third International Joint Conference on Artificial Intelligence,
  {IJCAI-24}}}, \bibfield{editor}{\bibinfo{person}{Kate Larson}} (Ed.).
  \bibinfo{publisher}{International Joint Conferences on Artificial
  Intelligence Organization}, \bibinfo{address}{Jeju, South Korea},
  \bibinfo{pages}{8048--8057}.
\newblock
\urldef\tempurl%
\url{https://doi.org/10.24963/ijcai.2024/890}
\showDOI{\tempurl}
\newblock
\shownote{Survey Track}.


\bibitem[Hong et~al\mbox{.}(2024)]%
        {Hong2024ArgMedAgentsEC}
\bibfield{author}{\bibinfo{person}{Shengxin Hong}, \bibinfo{person}{Liang
  Xiao}, \bibinfo{person}{Xin Zhang}, {and} \bibinfo{person}{Jianxia Chen}.}
  \bibinfo{year}{2024}\natexlab{}.
\newblock \showarticletitle{{ArgMed-Agents: Explainable Clinical Decision
  Reasoning with LLM Discussion via Argumentation Schemes}}. In
  \bibinfo{booktitle}{\emph{{Proceedings of the 2024 IEEE International
  Conference on Bioinformatics and Biomedicine (BIBM)}}}.
  \bibinfo{publisher}{IEEE}, \bibinfo{address}{Lisbon, Portugal},
  \bibinfo{pages}{5486--5493}.
\newblock


\bibitem[Ju et~al\mbox{.}(2024)]%
        {juFloodingSpreadManipulated2024}
\bibfield{author}{\bibinfo{person}{Tianjie Ju}, \bibinfo{person}{Yiting Wang},
  \bibinfo{person}{Xinbei Ma}, \bibinfo{person}{Pengzhou Cheng},
  \bibinfo{person}{Haodong Zhao}, \bibinfo{person}{Yulong Wang},
  \bibinfo{person}{Lifeng Liu}, \bibinfo{person}{Jian Xie},
  \bibinfo{person}{Zhuosheng Zhang}, {and} \bibinfo{person}{Gongshen Liu}.}
  \bibinfo{year}{2024}\natexlab{}.
\newblock \bibinfo{booktitle}{\emph{Flooding Spread of Manipulated Knowledge in
  LLM-Based Multi-Agent Communities}}.
\newblock School of Electronic Information and Electrical Engineering, Shanghai
  Jiao Tong University.
\newblock
\urldef\tempurl%
\url{https://doi.org/10.48550/arXiv.2407.07791}
\showDOI{\tempurl}
\showeprint[arXiv]{2407.07791}~[cs]


\bibitem[Kaesberg et~al\mbox{.}(2025)]%
        {kaesbergVotingConsensusDecisionMaking2025}
\bibfield{author}{\bibinfo{person}{Lars~Benedikt Kaesberg},
  \bibinfo{person}{Jonas Becker}, \bibinfo{person}{Jan~Philip Wahle},
  \bibinfo{person}{Terry Ruas}, {and} \bibinfo{person}{Bela Gipp}.}
  \bibinfo{year}{2025}\natexlab{}.
\newblock \bibinfo{booktitle}{\emph{Voting or Consensus? Decision-Making in
  Multi-Agent Debate}}.
\newblock University of Göttingen.
\newblock
\urldef\tempurl%
\url{https://doi.org/10.48550/arXiv.2502.19130}
\showDOI{\tempurl}
\showeprint[arXiv]{2502.19130}~[cs]


\bibitem[Li et~al\mbox{.}(2024b)]%
        {liSMoAImprovingMultiagent2024}
\bibfield{author}{\bibinfo{person}{Dawei Li}, \bibinfo{person}{Zhen Tan},
  \bibinfo{person}{Peijia Qian}, \bibinfo{person}{Yifan Li},
  \bibinfo{person}{Kumar~Satvik Chaudhary}, \bibinfo{person}{Lijie Hu}, {and}
  \bibinfo{person}{Jiayi Shen}.} \bibinfo{year}{2024}\natexlab{b}.
\newblock \bibinfo{booktitle}{\emph{SMoA: Improving Multi-agent Large Language
  Models with Sparse Mixture-of-Agents}}.
\newblock School of Computing, and Augmented Intelligence, Arizona State
  University.
\newblock
\urldef\tempurl%
\url{https://doi.org/10.48550/arXiv.2411.03284}
\showDOI{\tempurl}
\showeprint[arXiv]{2411.03284}~[cs]


\bibitem[Li et~al\mbox{.}(2024c)]%
        {liMoreAgentsAll2024}
\bibfield{author}{\bibinfo{person}{Junyou Li}, \bibinfo{person}{Qin Zhang},
  \bibinfo{person}{Yangbin Yu}, \bibinfo{person}{Qiang Fu}, {and}
  \bibinfo{person}{Deheng Ye}.} \bibinfo{year}{2024}\natexlab{c}.
\newblock \bibinfo{booktitle}{\emph{More Agents Is All You Need}}.
\newblock Tencent.
\newblock
\urldef\tempurl%
\url{https://doi.org/10.48550/arXiv.2402.05120}
\showDOI{\tempurl}
\showeprint[arXiv]{2402.05120}~[cs]


\bibitem[Li et~al\mbox{.}(2025)]%
        {li2025misfitting}
\bibfield{author}{\bibinfo{person}{Margaret Li}, \bibinfo{person}{Sneha
  Kudugunta}, {and} \bibinfo{person}{Luke Zettlemoyer}.}
  \bibinfo{year}{2025}\natexlab{}.
\newblock \bibinfo{title}{(Mis)Fitting: A Survey of Scaling Laws}.
\newblock
\newblock
\showeprint[arxiv]{2502.18969}~[cs.LG]
\urldef\tempurl%
\url{https://arxiv.org/abs/2502.18969}
\showURL{%
\tempurl}


\bibitem[Li et~al\mbox{.}(2024a)]%
        {liImprovingMultiAgentDebate2024}
\bibfield{author}{\bibinfo{person}{Yunxuan Li}, \bibinfo{person}{Yibing Du},
  \bibinfo{person}{Jiageng Zhang}, \bibinfo{person}{Le Hou},
  \bibinfo{person}{Peter Grabowski}, \bibinfo{person}{Yeqing Li}, {and}
  \bibinfo{person}{Eugene Ie}.} \bibinfo{year}{2024}\natexlab{a}.
\newblock \showarticletitle{Improving Multi-Agent Debate with Sparse
  Communication Topology}. In \bibinfo{booktitle}{\emph{Findings of the
  Association for Computational Linguistics: EMNLP 2024}},
  \bibfield{editor}{\bibinfo{person}{Yaser Al-Onaizan}, \bibinfo{person}{Mohit
  Bansal}, {and} \bibinfo{person}{Yun-Nung Chen}} (Eds.).
  \bibinfo{publisher}{Association for Computational Linguistics},
  \bibinfo{address}{Miami, Florida, USA}, \bibinfo{pages}{7281--7294}.
\newblock
\urldef\tempurl%
\url{https://doi.org/10.18653/v1/2024.findings-emnlp.427}
\showDOI{\tempurl}


\bibitem[Liang et~al\mbox{.}(2024)]%
        {Liang2023EncouragingDT}
\bibfield{author}{\bibinfo{person}{Tian Liang}, \bibinfo{person}{Zhiwei He},
  \bibinfo{person}{Wenxiang Jiao}, \bibinfo{person}{Xing Wang},
  \bibinfo{person}{Yan Wang}, \bibinfo{person}{Rui Wang},
  \bibinfo{person}{Yujiu Yang}, \bibinfo{person}{Shuming Shi}, {and}
  \bibinfo{person}{Zhaopeng Tu}.} \bibinfo{year}{2024}\natexlab{}.
\newblock \showarticletitle{{Encouraging Divergent Thinking in Large Language
  Models through Multi-Agent Debate}}. In
  \bibinfo{booktitle}{\emph{{Proceedings of the 2024 Conference on Empirical
  Methods in Natural Language Processing (EMNLP)}}}.
  \bibinfo{publisher}{Association for Computational Linguistics},
  \bibinfo{address}{Miami, FL, USA}, \bibinfo{pages}{17889--17904}.
\newblock


\bibitem[Liu et~al\mbox{.}(2024a)]%
        {liuGroupDebateEnhancingEfficiency2024}
\bibfield{author}{\bibinfo{person}{Tongxuan Liu}, \bibinfo{person}{Xingyu
  Wang}, \bibinfo{person}{Weizhe Huang}, \bibinfo{person}{Wenjiang Xu},
  \bibinfo{person}{Yuting Zeng}, \bibinfo{person}{Lei Jiang},
  \bibinfo{person}{Hailong Yang}, {and} \bibinfo{person}{Jing Li}.}
  \bibinfo{year}{2024}\natexlab{a}.
\newblock \bibinfo{booktitle}{\emph{GroupDebate: Enhancing the Efficiency of
  Multi-Agent Debate Using Group Discussion}}.
\newblock University of Science and Technology of China.
\newblock
\urldef\tempurl%
\url{https://doi.org/10.48550/arXiv.2409.14051}
\showDOI{\tempurl}
\showeprint[arXiv]{2409.14051}~[cs]


\bibitem[Liu et~al\mbox{.}(2024b)]%
        {liuAutonomousAgentsCollaborative2024}
\bibfield{author}{\bibinfo{person}{Wei Liu}, \bibinfo{person}{Chenxi Wang},
  \bibinfo{person}{Yifei Wang}, \bibinfo{person}{Zihao Xie},
  \bibinfo{person}{Rennai Qiu}, \bibinfo{person}{Yufan Dang},
  \bibinfo{person}{Zhuoyun Du}, \bibinfo{person}{Weize Chen},
  \bibinfo{person}{Cheng Yang}, {and} \bibinfo{person}{Chen Qian}.}
  \bibinfo{year}{2024}\natexlab{b}.
\newblock \bibinfo{title}{Autonomous Agents for Collaborative Task under
  Information Asymmetry}.
\newblock
\newblock
\showeprint[arxiv]{2406.14928}~[cs.AI]
\urldef\tempurl%
\url{https://arxiv.org/abs/2406.14928}
\showURL{%
\tempurl}


\bibitem[Liu et~al\mbox{.}(2024c)]%
        {liuDynamicLLMPoweredAgent2024}
\bibfield{author}{\bibinfo{person}{Zijun Liu}, \bibinfo{person}{Yanzhe Zhang},
  \bibinfo{person}{Peng Li}, \bibinfo{person}{Yang Liu}, {and}
  \bibinfo{person}{Diyi Yang}.} \bibinfo{year}{2024}\natexlab{c}.
\newblock \bibinfo{booktitle}{\emph{A Dynamic LLM-Powered Agent Network for
  Task-Oriented Agent Collaboration}}.
\newblock Dept. of Comp. Sci. \& Tech., Institute for AI, Tsinghua University,
  Beijing, China.
\newblock
\urldef\tempurl%
\url{https://doi.org/10.48550/arXiv.2310.02170}
\showDOI{\tempurl}
\showeprint[arXiv]{2310.02170}~[cs]


\bibitem[Maldonado et~al\mbox{.}(2024)]%
        {Multi-Agent-survey}
\bibfield{author}{\bibinfo{person}{Diego Maldonado}, \bibinfo{person}{Edison
  Cruz}, \bibinfo{person}{Jackeline Abad~Torres}, \bibinfo{person}{Patricio~J.
  Cruz}, {and} \bibinfo{person}{Silvana del~Pilar Gamboa~Benitez}.}
  \bibinfo{year}{2024}\natexlab{}.
\newblock \showarticletitle{Multi-Agent Systems: A Survey About Its Components,
  Framework and Workflow}.
\newblock \bibinfo{journal}{\emph{IEEE Access}}  \bibinfo{volume}{12}
  (\bibinfo{year}{2024}), \bibinfo{pages}{80950--80975}.
\newblock
\urldef\tempurl%
\url{https://doi.org/10.1109/ACCESS.2024.3409051}
\showDOI{\tempurl}


\bibitem[METR(2023)]%
        {post-training}
\bibfield{author}{\bibinfo{person}{METR}.} \bibinfo{year}{2023}\natexlab{}.
\newblock \bibinfo{title}{{M}easuring the impact of post-training enhancements
  --- metr.github.io}.
\newblock
  \bibinfo{howpublished}{\url{https://metr.github.io/autonomy-evals-guide/elicitation-gap//\#3.-results}}.
\newblock
\newblock
\shownote{[Accessed 25-11-2024]}.


\bibitem[Miller(2024)]%
        {millerAddingErrorBars2024}
\bibfield{author}{\bibinfo{person}{Evan Miller}.}
  \bibinfo{year}{2024}\natexlab{}.
\newblock \bibinfo{booktitle}{\emph{Adding Error Bars to Evals: A Statistical
  Approach to Language Model Evaluations}}.
\newblock Anthropic.
\newblock
\urldef\tempurl%
\url{https://doi.org/10.48550/arXiv.2411.00640}
\showDOI{\tempurl}
\showeprint[arXiv]{2411.00640}~[stat]


\bibitem[OpenAI(2025)]%
        {HttpsCdnopenaicomGpt45systemcard2272025pdf}
\bibfield{author}{\bibinfo{person}{OpenAI}.} \bibinfo{year}{2025}\natexlab{}.
\newblock \bibinfo{booktitle}{\emph{OpenAI GPT-4.5 System Card}}.
\newblock OpenAI.
\newblock
\urldef\tempurl%
\url{https://cdn.openai.com/gpt-4-5-system-card-2272025.pdf}
\showURL{%
\tempurl}


\bibitem[Pham et~al\mbox{.}(2024)]%
        {phamLetModelsSpeak2024}
\bibfield{author}{\bibinfo{person}{Chau Pham}, \bibinfo{person}{Boyi Liu},
  \bibinfo{person}{Yingxiang Yang}, \bibinfo{person}{Zhengyu Chen},
  \bibinfo{person}{Tianyi Liu}, \bibinfo{person}{Jianbo Yuan},
  \bibinfo{person}{Bryan~A. Plummer}, \bibinfo{person}{Zhaoran Wang}, {and}
  \bibinfo{person}{Hongxia Yang}.} \bibinfo{year}{2024}\natexlab{}.
\newblock \bibinfo{booktitle}{\emph{Let Models Speak Ciphers: Multiagent Debate
  through Embeddings}}.
\newblock Boston University.
\newblock
\urldef\tempurl%
\url{https://doi.org/10.48550/arXiv.2310.06272}
\showDOI{\tempurl}
\showeprint[arXiv]{2310.06272}~[cs]


\bibitem[Qian et~al\mbox{.}(2024)]%
        {Qian2023CommunicativeAF}
\bibfield{author}{\bibinfo{person}{Chen Qian}, \bibinfo{person}{Wei Liu},
  \bibinfo{person}{Hongzhang Liu}, \bibinfo{person}{Nuo Chen},
  \bibinfo{person}{Yufan Dang}, \bibinfo{person}{Jiahao Li},
  \bibinfo{person}{Cheng Yang}, \bibinfo{person}{Weize Chen},
  \bibinfo{person}{Yusheng Su}, \bibinfo{person}{Xin Cong},
  \bibinfo{person}{Juyuan Xu}, \bibinfo{person}{Dahai Li},
  \bibinfo{person}{Zhiyuan Liu}, {and} \bibinfo{person}{Maosong Sun}.}
  \bibinfo{year}{2024}\natexlab{}.
\newblock \bibinfo{title}{ChatDev: Communicative Agents for Software
  Development}.
\newblock
\newblock
\showeprint[arxiv]{2307.07924}~[cs.SE]
\urldef\tempurl%
\url{https://arxiv.org/abs/2307.07924}
\showURL{%
\tempurl}


\bibitem[Qian et~al\mbox{.}(2025)]%
        {qianScalingLargeLanguageModelbasedMultiAgent2025}
\bibfield{author}{\bibinfo{person}{Chen Qian}, \bibinfo{person}{Zihao Xie},
  \bibinfo{person}{YiFei Wang}, \bibinfo{person}{Wei Liu},
  \bibinfo{person}{Kunlun Zhu}, \bibinfo{person}{Hanchen Xia},
  \bibinfo{person}{Yufan Dang}, \bibinfo{person}{Zhuoyun Du},
  \bibinfo{person}{Weize Chen}, \bibinfo{person}{Cheng Yang},
  \bibinfo{person}{Zhiyuan Liu}, {and} \bibinfo{person}{Maosong Sun}.}
  \bibinfo{year}{2025}\natexlab{}.
\newblock \bibinfo{booktitle}{\emph{Scaling Large-Language-Model-based
  Multi-Agent Collaboration}}.
\newblock Tsinghua University.
\newblock
\urldef\tempurl%
\url{https://doi.org/10.48550/arXiv.2406.07155}
\showDOI{\tempurl}
\showeprint[arXiv]{2406.07155}~[cs]


\bibitem[Rasal and Hauer(2024)]%
        {rasalNavigatingComplexityOrchestrated2024}
\bibfield{author}{\bibinfo{person}{Sumedh Rasal} {and} \bibinfo{person}{E.~J.
  Hauer}.} \bibinfo{year}{2024}\natexlab{}.
\newblock \bibinfo{title}{Navigating Complexity: Orchestrated Problem Solving
  with Multi-Agent LLMs}.
\newblock
\newblock
\showeprint[arxiv]{2402.16713}~[cs.MA]
\urldef\tempurl%
\url{https://arxiv.org/abs/2402.16713}
\showURL{%
\tempurl}


\bibitem[Regan et~al\mbox{.}(2024)]%
        {reganProblemSolvingLanguageModel2024}
\bibfield{author}{\bibinfo{person}{Ciaran Regan}, \bibinfo{person}{Alexandre
  Gournail}, {and} \bibinfo{person}{Mizuki Oka}.}
  \bibinfo{year}{2024}\natexlab{}.
\newblock \bibinfo{booktitle}{\emph{Problem-Solving in Language Model
  Networks}}.
\newblock University of Tsukuba, Japan.
\newblock
\urldef\tempurl%
\url{https://doi.org/10.48550/arXiv.2406.12374}
\showDOI{\tempurl}
\showeprint[arXiv]{2406.12374}~[cs]


\bibitem[Rein et~al\mbox{.}(2023)]%
        {reinGPQAGraduateLevelGoogleProof2023}
\bibfield{author}{\bibinfo{person}{David Rein}, \bibinfo{person}{Betty~Li Hou},
  \bibinfo{person}{Asa~Cooper Stickland}, \bibinfo{person}{Jackson Petty},
  \bibinfo{person}{Richard~Yuanzhe Pang}, \bibinfo{person}{Julien Dirani},
  \bibinfo{person}{Julian Michael}, {and} \bibinfo{person}{Samuel~R. Bowman}.}
  \bibinfo{year}{2023}\natexlab{}.
\newblock \bibinfo{title}{{{GPQA}}: {{A Graduate-Level Google-Proof Q}}\&{{A
  Benchmark}}}.
\newblock
\newblock
\urldef\tempurl%
\url{https://doi.org/10.48550/arXiv.2311.12022}
\showDOI{\tempurl}
\showeprint[arxiv]{2311.12022}


\bibitem[Suzgun and Kalai(2024)]%
        {suzgunMetaPromptingEnhancingLanguage2024}
\bibfield{author}{\bibinfo{person}{Mirac Suzgun} {and}
  \bibinfo{person}{Adam~Tauman Kalai}.} \bibinfo{year}{2024}\natexlab{}.
\newblock \bibinfo{title}{Meta-Prompting: Enhancing Language Models with
  Task-Agnostic Scaffolding}.
\newblock
\newblock
\showeprint[arxiv]{2401.12954}~[cs.CL]
\urldef\tempurl%
\url{https://arxiv.org/abs/2401.12954}
\showURL{%
\tempurl}


\bibitem[Talebirad and Nadiri(2023)]%
        {Talebirad2023MultiAgentCH}
\bibfield{author}{\bibinfo{person}{Yashar Talebirad} {and}
  \bibinfo{person}{Amirhossein Nadiri}.} \bibinfo{year}{2023}\natexlab{}.
\newblock \bibinfo{title}{Multi-Agent Collaboration: Harnessing the Power of
  Intelligent LLM Agents}.
\newblock
\newblock
\showeprint[arxiv]{2306.03314}~[cs.AI]
\urldef\tempurl%
\url{https://arxiv.org/abs/2306.03314}
\showURL{%
\tempurl}


\bibitem[Wang et~al\mbox{.}(2023a)]%
        {wangLargeLanguageModels2023}
\bibfield{author}{\bibinfo{person}{Peiyi Wang}, \bibinfo{person}{Lei Li},
  \bibinfo{person}{Liang Chen}, \bibinfo{person}{Zefan Cai},
  \bibinfo{person}{Dawei Zhu}, \bibinfo{person}{Binghuai Lin},
  \bibinfo{person}{Yunbo Cao}, \bibinfo{person}{Qi Liu},
  \bibinfo{person}{Tianyu Liu}, {and} \bibinfo{person}{Zhifang Sui}.}
  \bibinfo{year}{2023}\natexlab{a}.
\newblock \bibinfo{booktitle}{\emph{Large {{Language Models}} Are Not {{Fair
  Evaluators}}}}.
\newblock National Key Laboratory for Multimedia Information Processing, Peking
  University.
\newblock
\urldef\tempurl%
\url{https://doi.org/10.48550/arXiv.2305.17926}
\showDOI{\tempurl}
\showeprint[arXiv]{2305.17926}~[cs]


\bibitem[Wang et~al\mbox{.}(2023b)]%
        {Wang2023OnTD}
\bibfield{author}{\bibinfo{person}{Qineng Wang}, \bibinfo{person}{Zihao Wang},
  \bibinfo{person}{Ying Su}, {and} \bibinfo{person}{Yangqiu Song}.}
  \bibinfo{year}{2023}\natexlab{b}.
\newblock \bibinfo{title}{On the Discussion of Large Language Models: Symmetry
  of Agents and Interplay with Prompts}.
\newblock
\newblock
\showeprint[arxiv]{2311.07076}~[cs.CL]
\urldef\tempurl%
\url{https://arxiv.org/abs/2311.07076}
\showURL{%
\tempurl}


\bibitem[Wei et~al\mbox{.}(2023)]%
        {wei2023chainofthoughtpromptingelicitsreasoning}
\bibfield{author}{\bibinfo{person}{Jason Wei}, \bibinfo{person}{Xuezhi Wang},
  \bibinfo{person}{Dale Schuurmans}, \bibinfo{person}{Maarten Bosma},
  \bibinfo{person}{Brian Ichter}, \bibinfo{person}{Fei Xia},
  \bibinfo{person}{Ed Chi}, \bibinfo{person}{Quoc Le}, {and}
  \bibinfo{person}{Denny Zhou}.} \bibinfo{year}{2023}\natexlab{}.
\newblock \bibinfo{title}{Chain-of-Thought Prompting Elicits Reasoning in Large
  Language Models}.
\newblock
\newblock
\showeprint[arxiv]{2201.11903}~[cs.CL]
\urldef\tempurl%
\url{https://arxiv.org/abs/2201.11903}
\showURL{%
\tempurl}


\bibitem[Wu et~al\mbox{.}(2023)]%
        {Wu2023LargeLM}
\bibfield{author}{\bibinfo{person}{Ning Wu}, \bibinfo{person}{Ming Gong},
  \bibinfo{person}{Linjun Shou}, \bibinfo{person}{Shining Liang}, {and}
  \bibinfo{person}{Daxin Jiang}.} \bibinfo{year}{2023}\natexlab{}.
\newblock \bibinfo{title}{Large Language Models are Diverse Role-Players for
  Summarization Evaluation}.
\newblock
\newblock
\showeprint[arxiv]{2303.15078}~[cs.CL]
\urldef\tempurl%
\url{https://arxiv.org/abs/2303.15078}
\showURL{%
\tempurl}


\bibitem[Xiong et~al\mbox{.}(2023)]%
        {Xiong2023DivingIT}
\bibfield{author}{\bibinfo{person}{Kai Xiong}, \bibinfo{person}{Xiao Ding},
  \bibinfo{person}{Yixin Cao}, \bibinfo{person}{Ting Liu}, {and}
  \bibinfo{person}{Bing Qin}.} \bibinfo{year}{2023}\natexlab{}.
\newblock \showarticletitle{Examining Inter-Consistency of Large Language
  Models Collaboration: An In-depth Analysis via Debate}. In
  \bibinfo{booktitle}{\emph{Findings of the Association for Computational
  Linguistics: EMNLP 2023}}, \bibfield{editor}{\bibinfo{person}{Houda Bouamor},
  \bibinfo{person}{Juan Pino}, {and} \bibinfo{person}{Kalika Bali}} (Eds.).
  \bibinfo{publisher}{Association for Computational Linguistics},
  \bibinfo{address}{Singapore}, \bibinfo{pages}{7572--7590}.
\newblock
\urldef\tempurl%
\url{https://doi.org/10.18653/v1/2023.findings-emnlp.508}
\showDOI{\tempurl}


\bibitem[Xu et~al\mbox{.}(2023)]%
        {xuTowardsReasoningLargeLanguage2023}
\bibfield{author}{\bibinfo{person}{Zhenran Xu}, \bibinfo{person}{Senbao Shi},
  \bibinfo{person}{Baotian Hu}, \bibinfo{person}{Jindi Yu},
  \bibinfo{person}{Dongfang Li}, \bibinfo{person}{Min Zhang}, {and}
  \bibinfo{person}{Yuxiang Wu}.} \bibinfo{year}{2023}\natexlab{}.
\newblock \bibinfo{title}{Towards Reasoning in Large Language Models via
  Multi-Agent Peer Review Collaboration}.
\newblock
\newblock
\showeprint[arxiv]{2311.08152}~[cs.CL]
\urldef\tempurl%
\url{https://arxiv.org/abs/2311.08152}
\showURL{%
\tempurl}


\bibitem[Ye et~al\mbox{.}(2024)]%
        {yeMultiAgentSamplingScaling2024}
\bibfield{author}{\bibinfo{person}{Hai Ye}, \bibinfo{person}{Mingbao Lin},
  \bibinfo{person}{Hwee~Tou Ng}, {and} \bibinfo{person}{Shuicheng Yan}.}
  \bibinfo{year}{2024}\natexlab{}.
\newblock \bibinfo{booktitle}{\emph{Multi-Agent Sampling: Scaling Inference
  Compute for Data Synthesis with Tree Search-Based Agentic Collaboration}}.
\newblock National University of Singapore.
\newblock
\urldef\tempurl%
\url{https://doi.org/10.48550/arXiv.2412.17061}
\showDOI{\tempurl}
\showeprint[arXiv]{2412.17061}~[cs]


\bibitem[Yin et~al\mbox{.}(2023)]%
        {yinExchangeThoughtEnhancingLarge2023}
\bibfield{author}{\bibinfo{person}{Zhangyue Yin}, \bibinfo{person}{Qiushi Sun},
  \bibinfo{person}{Cheng Chang}, \bibinfo{person}{Qipeng Guo},
  \bibinfo{person}{Junqi Dai}, \bibinfo{person}{Xuanjing Huang}, {and}
  \bibinfo{person}{Xipeng Qiu}.} \bibinfo{year}{2023}\natexlab{}.
\newblock \showarticletitle{Exchange-of-{{Thought}}: {{Enhancing Large Language
  Model Capabilities}} through {{Cross-Model Communication}}}. In
  \bibinfo{booktitle}{\emph{Proceedings of the 2023 {{Conference}} on
  {{Empirical Methods}} in {{Natural Language Processing}}}},
  \bibfield{editor}{\bibinfo{person}{Houda Bouamor}, \bibinfo{person}{Juan
  Pino}, {and} \bibinfo{person}{Kalika Bali}} (Eds.).
  \bibinfo{publisher}{Association for Computational Linguistics},
  \bibinfo{address}{Singapore}, \bibinfo{pages}{15135--15153}.
\newblock
\urldef\tempurl%
\url{https://doi.org/10.18653/v1/2023.emnlp-main.936}
\showDOI{\tempurl}


\bibitem[Zhang et~al\mbox{.}(2024b)]%
        {zhangCutCrapEconomical2024}
\bibfield{author}{\bibinfo{person}{Guibin Zhang}, \bibinfo{person}{Yanwei Yue},
  \bibinfo{person}{Zhixun Li}, \bibinfo{person}{Sukwon Yun},
  \bibinfo{person}{Guancheng Wan}, \bibinfo{person}{Kun Wang},
  \bibinfo{person}{Dawei Cheng}, \bibinfo{person}{Jeffrey~Xu Yu}, {and}
  \bibinfo{person}{Tianlong Chen}.} \bibinfo{year}{2024}\natexlab{b}.
\newblock \bibinfo{booktitle}{\emph{Cut the Crap: An Economical Communication
  Pipeline for LLM-based Multi-Agent Systems}}.
\newblock Tongji University.
\newblock
\urldef\tempurl%
\url{https://doi.org/10.48550/arXiv.2410.02506}
\showDOI{\tempurl}
\showeprint[arXiv]{2410.02506}~[cs]


\bibitem[Zhang et~al\mbox{.}(2025)]%
        {zhangGDesignerArchitectingMultiagent2025}
\bibfield{author}{\bibinfo{person}{Guibin Zhang}, \bibinfo{person}{Yanwei Yue},
  \bibinfo{person}{Xiangguo Sun}, \bibinfo{person}{Guancheng Wan},
  \bibinfo{person}{Miao Yu}, \bibinfo{person}{Junfeng Fang},
  \bibinfo{person}{Kun Wang}, \bibinfo{person}{Tianlong Chen}, {and}
  \bibinfo{person}{Dawei Cheng}.} \bibinfo{year}{2025}\natexlab{}.
\newblock \bibinfo{booktitle}{\emph{G-Designer: Architecting Multi-agent
  Communication Topologies via Graph Neural Networks}}.
\newblock CUHK.
\newblock
\urldef\tempurl%
\url{https://doi.org/10.48550/arXiv.2410.11782}
\showDOI{\tempurl}
\showeprint[arXiv]{2410.11782}~[cs]


\bibitem[Zhang et~al\mbox{.}(2024a)]%
        {zhang2024exploringcollaborationmechanismsllm}
\bibfield{author}{\bibinfo{person}{Jintian Zhang}, \bibinfo{person}{Xin Xu},
  \bibinfo{person}{Ningyu Zhang}, \bibinfo{person}{Ruibo Liu},
  \bibinfo{person}{Bryan Hooi}, {and} \bibinfo{person}{Shumin Deng}.}
  \bibinfo{year}{2024}\natexlab{a}.
\newblock \bibinfo{title}{Exploring Collaboration Mechanisms for LLM Agents: A
  Social Psychology View}.
\newblock
\newblock
\showeprint[arxiv]{2310.02124}~[cs.CL]
\urldef\tempurl%
\url{https://arxiv.org/abs/2310.02124}
\showURL{%
\tempurl}


\end{thebibliography}

\appendix

\end{document}